\documentclass[final,12pt,onecolumn]{elsarticle} 

 \usepackage{graphicx}
 \usepackage{epsfig}
\usepackage{amssymb}
\usepackage{amsthm}
\usepackage{amsmath}
\usepackage{multirow}
\usepackage{lineno}
\usepackage{subfigure}

\journal{Astroparticle Physics}

\begin{document}
\begin{frontmatter}



\title{Searches for small-scale anisotropies from neutrino point sources with three years of 
IceCube data}



\author[Adelaide]{M.~G.~Aartsen} 
\author[Zeuthen]{M.~Ackermann} 
\author[Christchurch]{J.~Adams} 
\author[Geneva]{J.~A.~Aguilar} 
\author[MadisonPAC]{M.~Ahlers} 
\author[StockholmOKC]{M.~Ahrens} 
\author[Erlangen]{D.~Altmann} 
\author[PennPhys]{T.~Anderson} 
\author[MadisonPAC]{C.~Arguelles} 
\author[PennPhys]{T.~C.~Arlen} 
% ...
\address[Adelaide]{School of Chemistry \& Physics, University of Adelaide, Adelaide SA, 5005 Australia}
% ...
\address[Christchurch]{Dept.~of Physics and Astronomy, University of Canterbury, Private Bag 4800, Christchurch, New Zealand}
% ...
\address[Erlangen]{Erlangen Centre for Astroparticle Physics, Friedrich-Alexander-Universit\"at Erlangen-N\"urnberg, D-91058 Erlangen, Germany}
\address[Geneva]{D\'epartement de physique nucl\'eaire et corpusculaire, Universit\'e de Gen\`eve, CH-1211 Gen\`eve, Switzerland}
% ...
\address[MadisonPAC]{Dept.~of Physics and Wisconsin IceCube Particle Astrophysics Center, University of Wisconsin, Madison, WI 53706, USA}
% ...
\address[StockholmOKC]{Oskar Klein Centre and Dept.~of Physics, Stockholm University, SE-10691 Stockholm, Sweden}
% ...
\address[PennPhys]{Dept.~of Physics, Pennsylvania State University, University Park, PA 16802, USA}
% ...
\address[Zeuthen]{DESY, D-15735 Zeuthen, Germany}
\fntext[Tokyofn]{Earthquake Research Institute, University of Tokyo, Bunkyo, Tokyo 113-0032, Japan}
\fntext[Goddard]{NASA Goddard Space Flight Center, Greenbelt, MD 20771, USA}
\cortext[cor1]{Corresponding authors: \hspace*{2mm} Anna.Bernhard@tum.de, \\
\noindent\hspace*{46mm} Leuermann@physik.rwth-aachen.de}

\newpage

\begin{abstract}

Recently, IceCube found evidence for a diffuse signal of astrophysical 
neutrinos in an energy range of $\sim 60\,\mathrm{TeV}$ to the $\mathrm{PeV}$-scale~\cite{HESE_2year}. The origin of those events, being a key to understanding the origin of cosmic rays, is still an unsolved question. So far, analyses have not succeeded to resolve the 
diffuse signal into point-like sources. Searches including a maximum-likelihood-ratio test, based on the reconstructed directions and energies of the detected down- and up-going neutrino candidates, were also performed on IceCube 
data leading to the exclusion of bright point sources. In this paper, we present two methods to search for faint
neutrino point sources in three years of IceCube data, taken between 2008 and 2011. The first 
method is an autocorrelation test,  applied separately to the northern 
and southern sky. The second method is a multipole analysis, which 
expands the measured data in the northern hemisphere into spherical harmonics and uses the 
resulting expansion coefficients to separate signal from background. 
 With both methods, the 
results are consistent with the background expectation with a slightly more sparse spatial distribution, corresponding to an underfluctuation. Depending on the assumed number of sources, the resulting upper limit on the flux per source in the northern hemisphere for an $E^{-2}$ energy spectrum ranges from $\sim 1.5 \cdot 10^{-8}$\,GeV/$\mathrm{cm}^{2}$ $\mathrm{s}^{-1}$, in the case of one assumed source, to $\sim 4 \cdot 10^{-10}$\,GeV/$\mathrm{cm}^{2}$ $\mathrm{s}^{-1}$, in the case of 3500 assumed sources.

\end{abstract}


\end{frontmatter}

\newpage
\clearpage


\section{Introduction}
\label{}

The unsolved problem of the origin of cosmic rays is one of the 
biggest challenges in high-energy astrophysics. In hadronic 
interactions of cosmic rays with matter, high-energy neutrinos are produced in the direct environment of cosmic ray sources. Possible candidates for these sources are, for example, Gamma Ray Bursts (GRBs)~\cite{Mesz_GRB, Waxman_Nu} or Active galactic Nuclei (AGNs)~\cite{HE_AGNs, WaxBa_nu}. From Fermi Acceleration, the resulting energy-dependent differential flux ($\frac{\mathrm{d}\phi}{\mathrm{d}E_\nu}$) of astrophysical neutrinos is expected to follow a power law of $E_\nu^{-\gamma}$, where $E_\nu$ is the neutrino energy and $\gamma = 2+\epsilon$ is close to the spectral index of the cosmic ray production process~\cite{FrancisAGNs}. The so-called inefficiency ($\epsilon$) is often assumed to be  $0\leq \epsilon \leq 1$, while $\gamma \approx 2$ is favored by Fermi Acceleration~\cite{FrancisAGNs}.\\
Neutrinos are also expected in coincidence with high-energy photons.  Thus, the measurement of thousands of gamma-ray sources, like AGNs~\cite{AGNGammaToNu, FermiAGNs}, provides a large number of potential neutrino sources. \\
Since neutrinos do not experience deflections or scattering, 
they are ideal messengers for observing and tracing the hadronic 
interactions described above. The detection of high-energy neutrinos of cosmic origin would give important insights for identifying the sources of cosmic rays~\cite{Tjus_sources, WaxBah}. \\
The IceCube neutrino telescope~\cite{icecube} recently detected a diffuse cosmic flux of high-energy neutrino events in two years of IceCube data by searching for neutrino-induced events
with an interaction vertex within the detector~\cite{HESE_2year}. This permits one to reduce the dominant atmospheric muon background. In a follow-up analysis, this diffuse flux was investigated in more detail using three years of IceCube data~\cite{HESE}. This high energy starting events analysis (HESE) yielded 37 events compared to an expected background of 15.0 events from atmospheric muons and neutrinos. These events could provide information about potential cosmic ray astrophysical sources and motivate additional searches. 

IceCube, with its 
large field-of-view, offers a unique opportunity to study the 
production and interaction of high-energy cosmic rays using neutrinos. 
The detector, which is located at the geographical South Pole, has a detection volume of $\sim1\,\mathrm{km}^3$ deep in the Antarctic glacier and 
an additional $\sim1\,\mathrm{km}^2$ surface air shower detector, called IceTop. The IceCube detector consists of digital optical modules~(DOMs)~\cite{DAQ}, placed on strings deployed vertically at depths between 1450\,m and 2450\,m. The strings hold 60~DOMs each equipped with a photomultiplier tube and digitizing electronics  to detect neutrinos by measuring Cherenkov radiation of their secondary particles~\cite{PMTs}. A detailed description of the data acquisition system can be found in~\cite{DAQ}. The detector was built in several stages between 2005 and 2010, such that each year several strings were 
added, until reaching the final configuration of 86-strings, containing more than 5000~DOMs. 
IceCube has completed different point source searches, including an energy-dependent likelihood point 
source search scanning the full sky~\cite{all_sky}, as well as searches for flaring and periodic neutrino emission~\cite{flare}. Additionally there are searches for diffuse neutrino 
emission looking for deviations in the two dimensional 
distribution of energy and zenith angle~\cite{diffuse}.
Point source searches are most sensitive for finding individual sources of astrophysical neutrinos among the background of atmospheric events (neutrinos and muons from cosmic ray interactions at Earth).  Diffuse searches, on the other hand, are most sensitive for detecting within this background the presence of high-energy astrophysical neutrinos throughout the sky, without identifying individual sources.  In between these two scenarios is the possibility that many weak sources exist.  These could contribute to the detected diffuse signal and create a small number of events clustering on the background event distribution, while the individual clusters remain too weak to be detected by the point source searches.  In this paper we present two searches for such small-scale clustering.

The first is an autocorrelation test using a two-point 
(2-pt) correlation function performed in the northern and southern hemisphere. Since most of the signal-like events are at high energies, we extended the autocorrelation test to include the most energetic events in an additional test (2-pt HE). The second test is a multipole expansion of the skymap of neutrino arrival 
directions. The goal of both methods is to gain sensitivity to 
faint sources at unknown positions in the sky with unidentified 
energy spectra using a three year data-set of the partially completed detector. Both methods are 
complementary to previous searches and therefore an 
important addition to IceCube searches for a cosmic neutrino flux. 
A similar search for clustering was recently performed by ANTARES, details can be found in \cite{Antares_2pt}.

The paper is organized as follows: In section~\ref{datasample}, the 
data sample used and the generation of pseudo experiments is 
described. The 2-pt correlation test is explained in 
section~\ref{sec:method_2pt}, and the multipole analysis in 
section~\ref{sec:method_mp}. The performance of both analyses, decribed by the 
discovery potential, is given in 
section~\ref{sec:DataAnalysis}. In section~\ref{sec:Result}, the 
experimental result is presented and exclusion limits are calculated.
Systematic uncertainties are discussed in section~\ref{sec:systematics}. 
In section~\ref{sec:conclusion} a conclusion is drawn. 


\section{Data Sample}
\label{datasample}

\subsection{Experimental Data}

The data presented was taken between 2008 
and 2011. It consists of 3 different detector configurations: 
40-, 59- and 79-string configurations (IC40, IC59 and IC79, respectively), where the labeling corresponds to the number of deployed strings. For the selection criteria applied~\cite{all_sky}, 108\,310 and 146\,047 track-like events were found in the northern and southern sky, respectively.
%
The data are background dominated, consisting mostly of atmospheric muons in the southern sky and of atmospheric neutrinos in the northern sky where the atmospheric muon background is shielded by the Earth. At trigger level, there is a significant contribution from atmospheric muons also in the northern sky, caused by mis-reconstructed atmospheric muons near the horizon and events with two muons within the same readout window reconstructed as a single track~\cite{EnergyReco}. This background is reduced by rejecting events with a poor angular resolution and events with multiple muon tracks which are identified by a clustering algorithm. Requirements on the amount of light seen in the detector are used to also reduce the atmospheric neutrino background in the northern hemisphere and result in a neutrino energy threshold of around 100 GeV. Following these guidelines, the event selection was performed separately for each detector configuration. For IC40, straight cuts were used~\cite{IC40_PS, all_sky}, while for IC59 and IC79, the event selection was performed using several Boosted Decision Trees~\cite{all_sky}. To reduce the atmospheric muon background in the southern hemisphere, declination dependent cuts on the reconstructed muon energy were developed for each data sample. 
While the cuts were designed to result in a smooth distribution of the events, there are remaining disuniformities induced by the steep change of the event density with the declination if a smaller binning is used to present the data than was used in the event selection. The effects of these disuniformities are fully included in the background estimation and the calculated statistical significance of the result.
For each selected event, the declination, the right ascension and the energy of the muon inside the detector were reconstructed, using the methods described in \cite{muontrack} and \cite{EnergyReco}. Atmospheric neutrinos yield events with reconstructed muon energies of up to $\sim$100 TeV. Above this energy, their flux is too low to make a significant contribution. The atmospheric muon background in the southern hemisphere however is significantly larger and contributes events with up to 10 times higher energies. The declination distribution of the experimental data after the event selection is shown in Fig. 1(a).

\subsection{Signal Simulation}\label{sec:datasample_chara}
\begin{figure*}[ht]
 \centering \subfigure[Experimental declination distribution]{
 \includegraphics[width=0.45\textwidth]
 {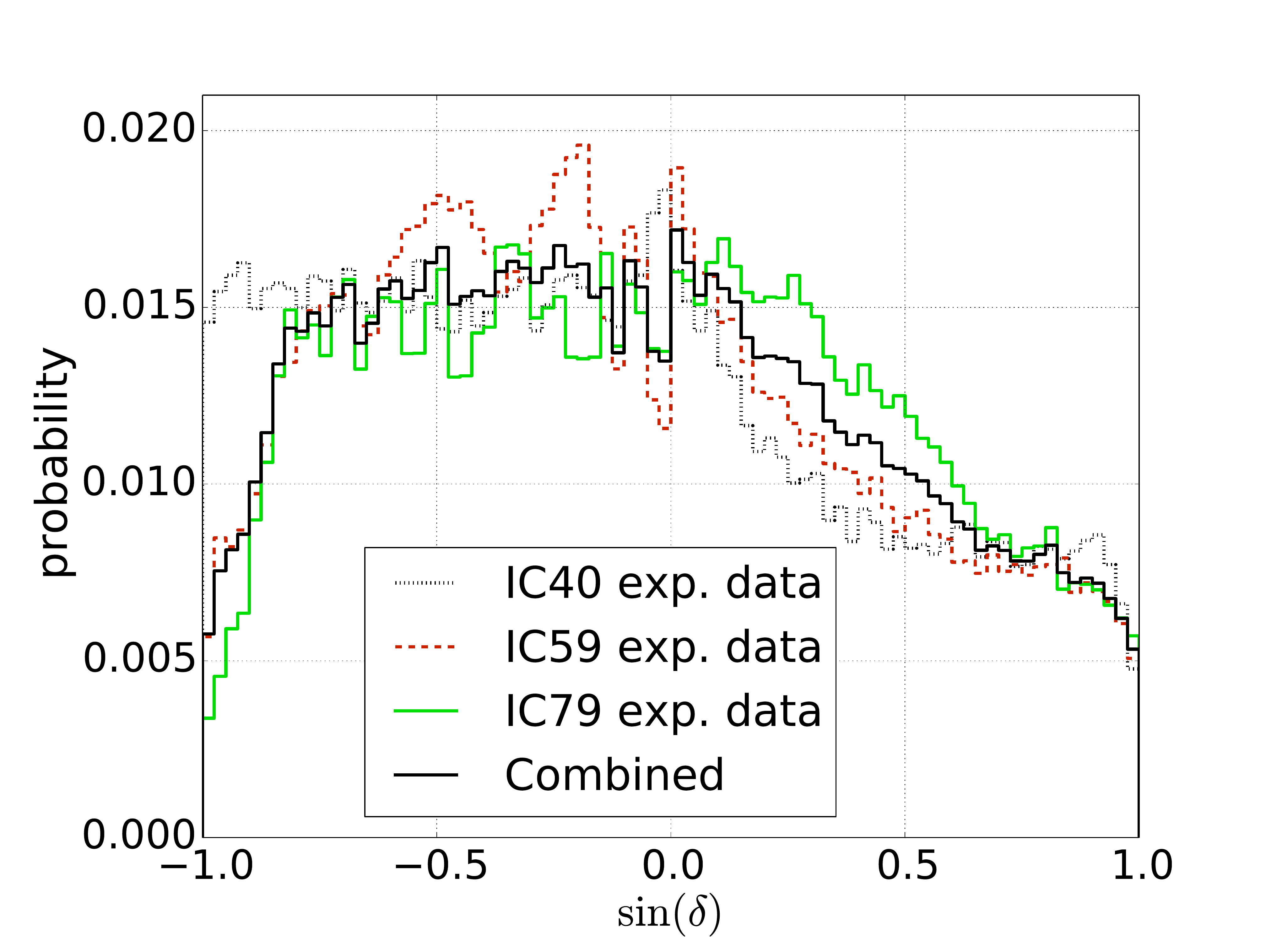} \label{fig:expZenith} } 
 \subfigure[$E^{-2}$ Point Spread Function from MC]{\includegraphics
 [width=0.45\textwidth]{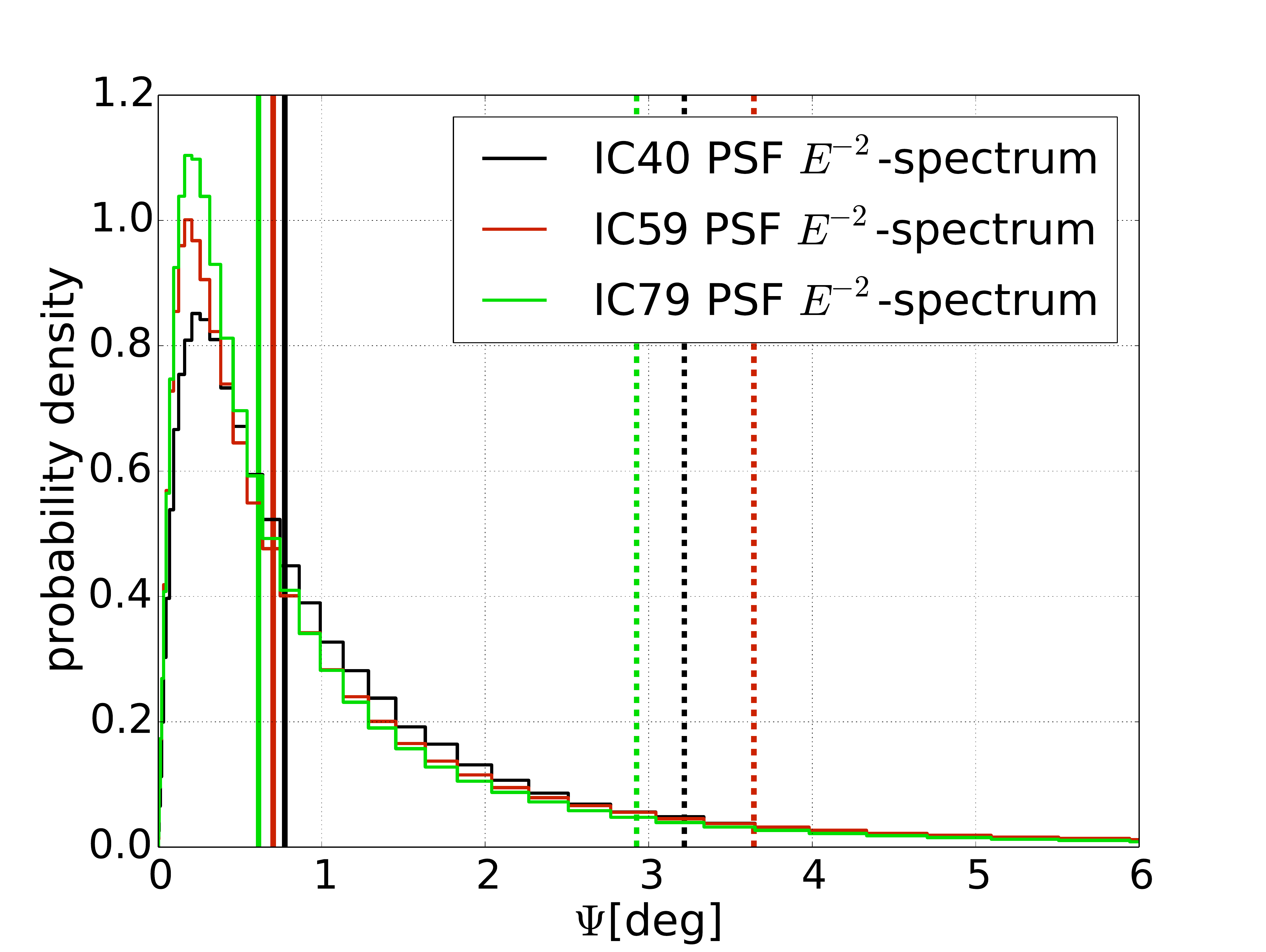} \label{fig:PSFs} } 
 \caption{(a) Experimental declination distribution and (b) Point Spread 
 Function (PSF) obtained from full detector MC simulations. In (b) the median and the 90\%-quantile for each distribution 
 are shown as solid and dashed vertical lines, respectively.}
\end{figure*}

The energy spectrum of candidate neutrinos is motivated by Fermi acceleration~\cite{FrancisAGNs}, and is assumed to follow a power law of $E^{-\gamma}$ with different spectral indices ($\gamma$). Additionally, the number of sources ($N_{\mathrm{Sou}}$) is varied between one and several thousand sources with respect to the wide range of possible source candidates. From full detector Monte-Carlo simulation (MC)~\cite{MC}, the detection efficiency and the angular resolution of these signal events can be estimated. Since the angular 
resolution of the detector depends on the  energy spectrum, the 
distribution of the angular reconstruction error, i.e. the Point Spread Function~(PSF), is obtained for each of the investigated energy spectra ($E^{-2}$, 
$E^{-2.25}$ and $E^{-3}$) and for each detector configuration. As an example, for an 
$E^{-2}$ spectrum, the PSFs for IC40, IC59 and IC79 are
shown in Fig.~\ref{fig:PSFs}.

\begin{figure}[tb]
     \centering
     \subfigure[Northern hemisphere]{
     \includegraphics[width=0.48\textwidth]{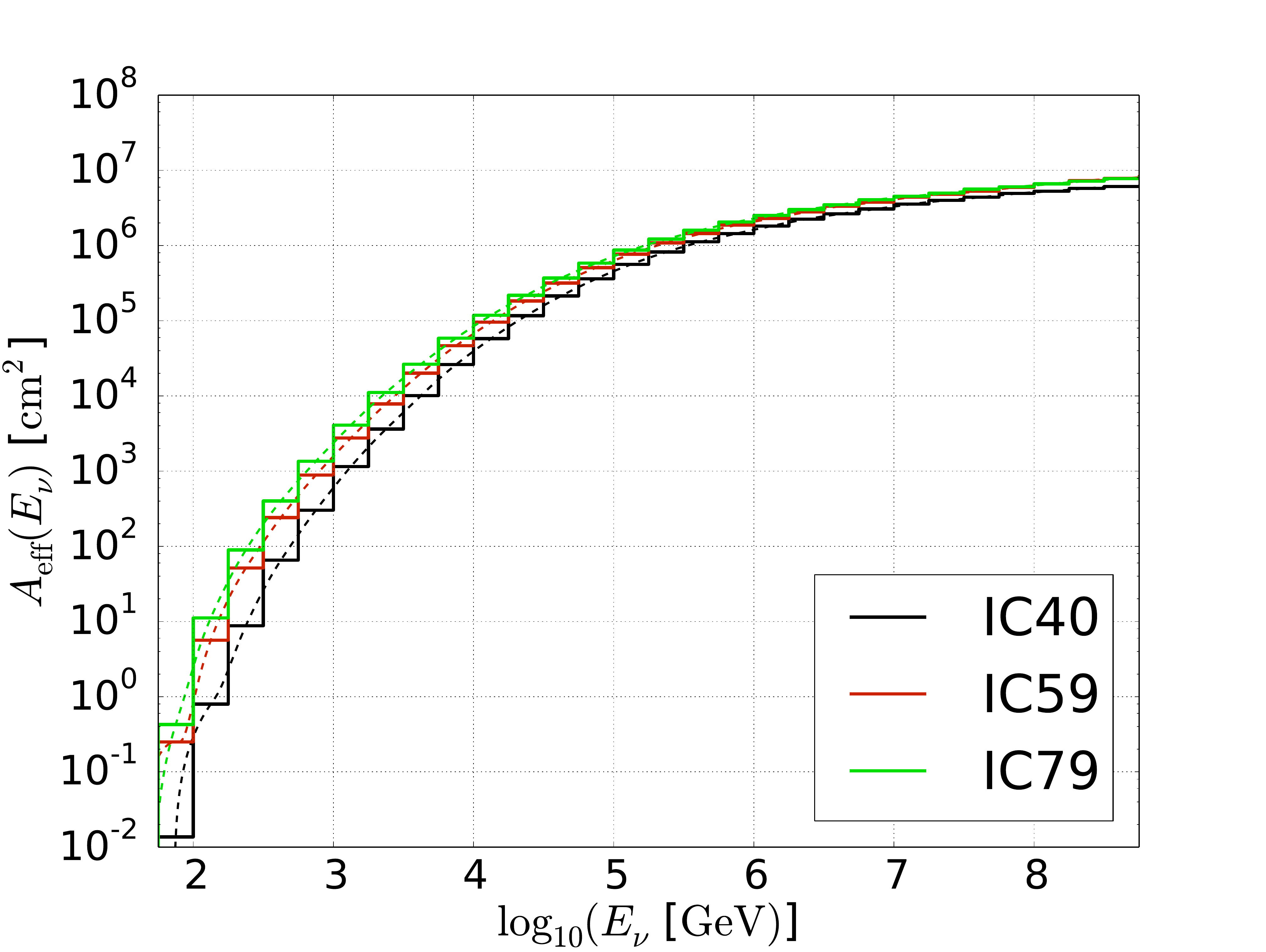}}
     \subfigure[Southern hemisphere]{
     \includegraphics[width=0.48\textwidth]{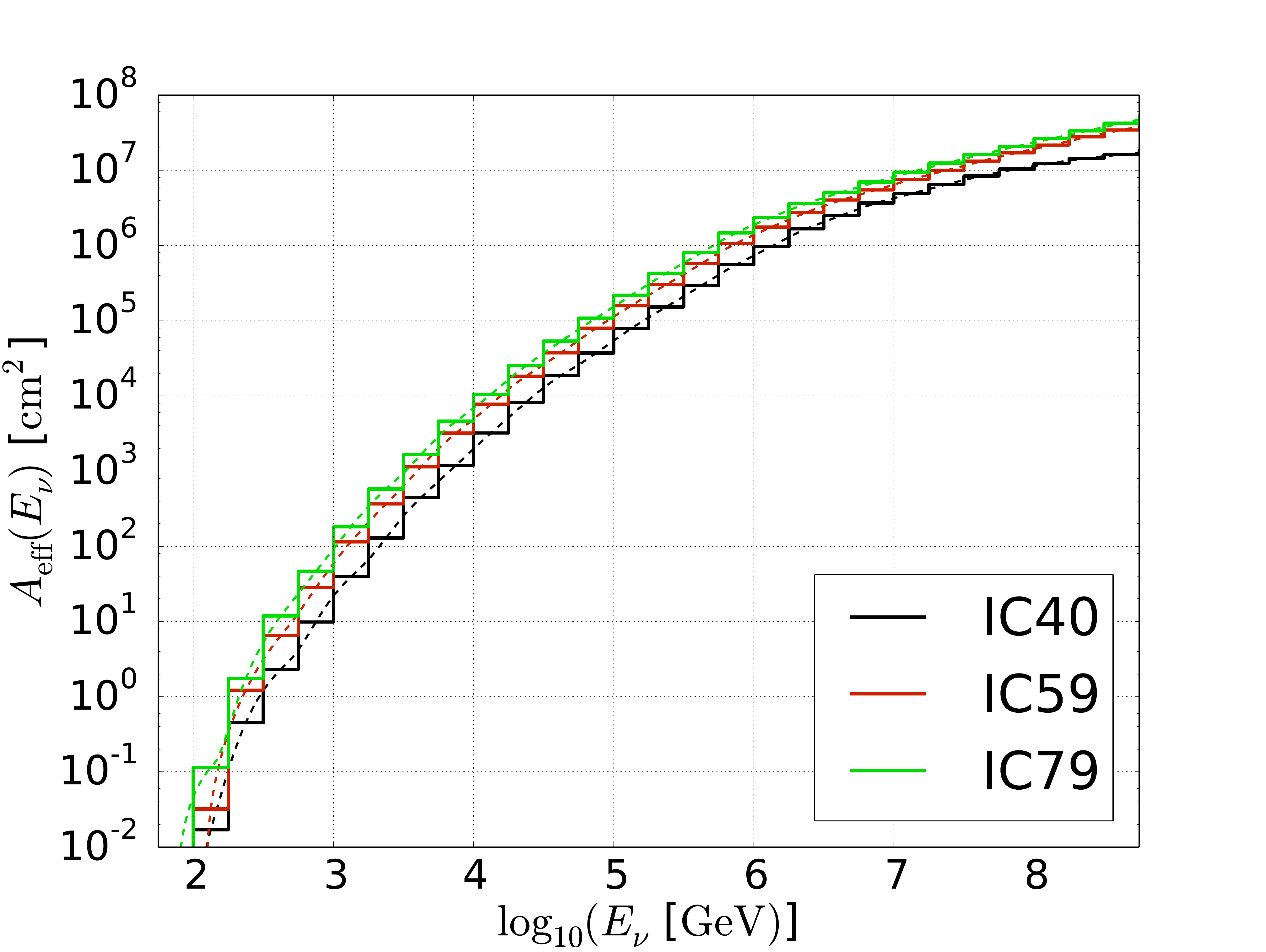}}
     \caption{The solid-angle-averaged effective area of IC40, IC59, and IC79 for neutrinos from the northern hemisphere (a) and the southern hemisphere (b).  The effective area is determined from Monte Carlo simulations for the final event selection and is a function of the neutrino energy $E_{\nu}$ and the incident zenith angle (not shown here, but averaged over solid angle).}
     \label{effArea}
\end{figure}

The detector efficiency is a function of the energy E and the zenith angle $\theta$ of an incoming neutrino. It can be characterized by the effective area $A_{\mathrm{eff}}(E, \theta)$, which is the corresponding area of a hypothetical detector, that has 100\,\% efficiency for detecting neutrinos generated uniformly and omnidirectionally at the surface of the atmosphere of Earth with given energy $E$ and zenith angle $\theta$.  For a neutrino source at a given declination in the sky, the number of detected neutrinos is proportional to a convolution of the energy-dependent effective area (evaluated at the appropriate zenith angle) and the energy spectrum of the source.  As an illustration, the solid-angle-averaged effective area $A_{\mathrm{eff}}(E)$ for the northern and southern sky are shown in Fig.~\ref{effArea} for the final event selection.

\begin{figure*}[tb]
 \centering 
 \subfigure[For different energy spectra]{\includegraphics[width=0.48\textwidth]
 {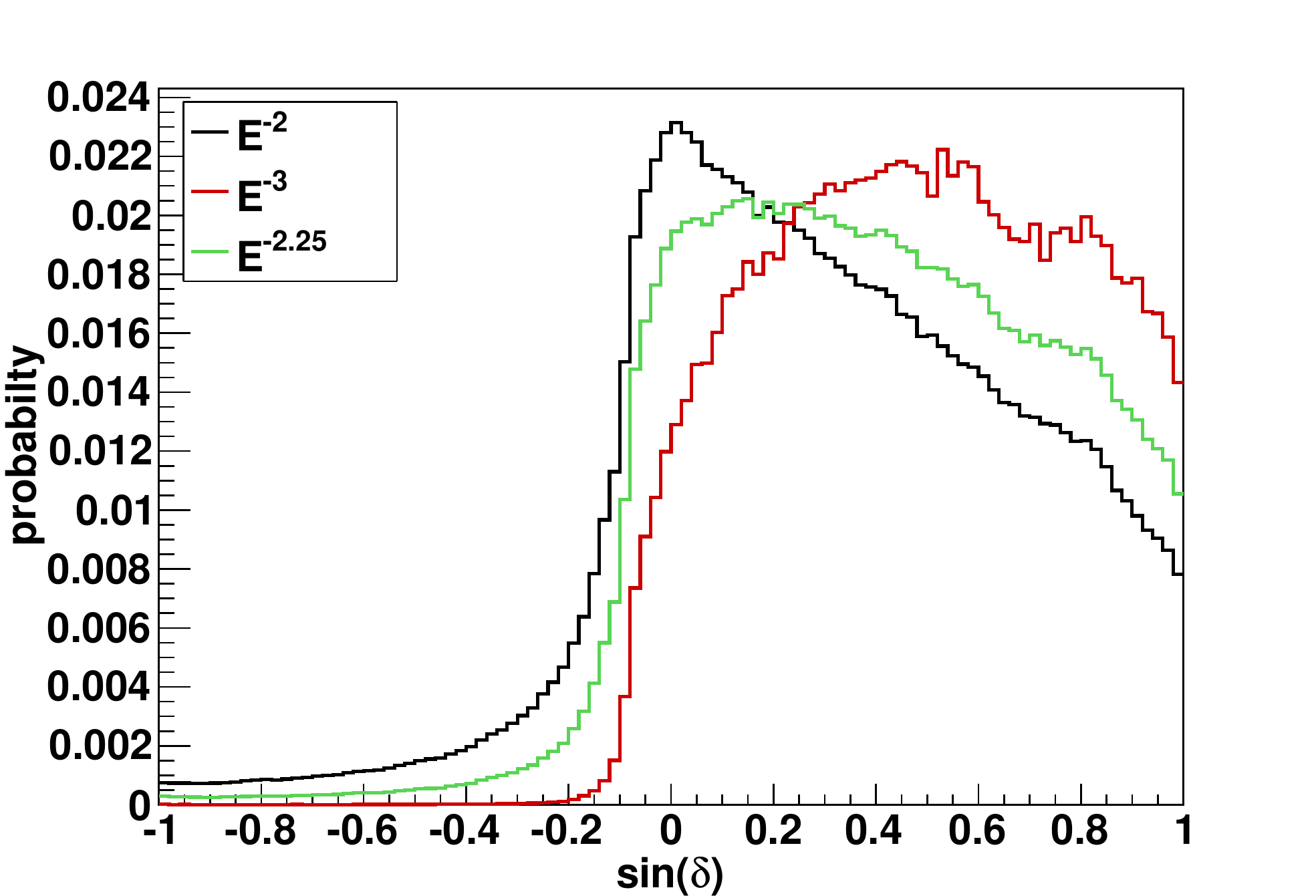} \label{fig:sigZenith} } 
 \subfigure[For different energy bins of $E^{-2}$]{\includegraphics[width=0.48\textwidth]{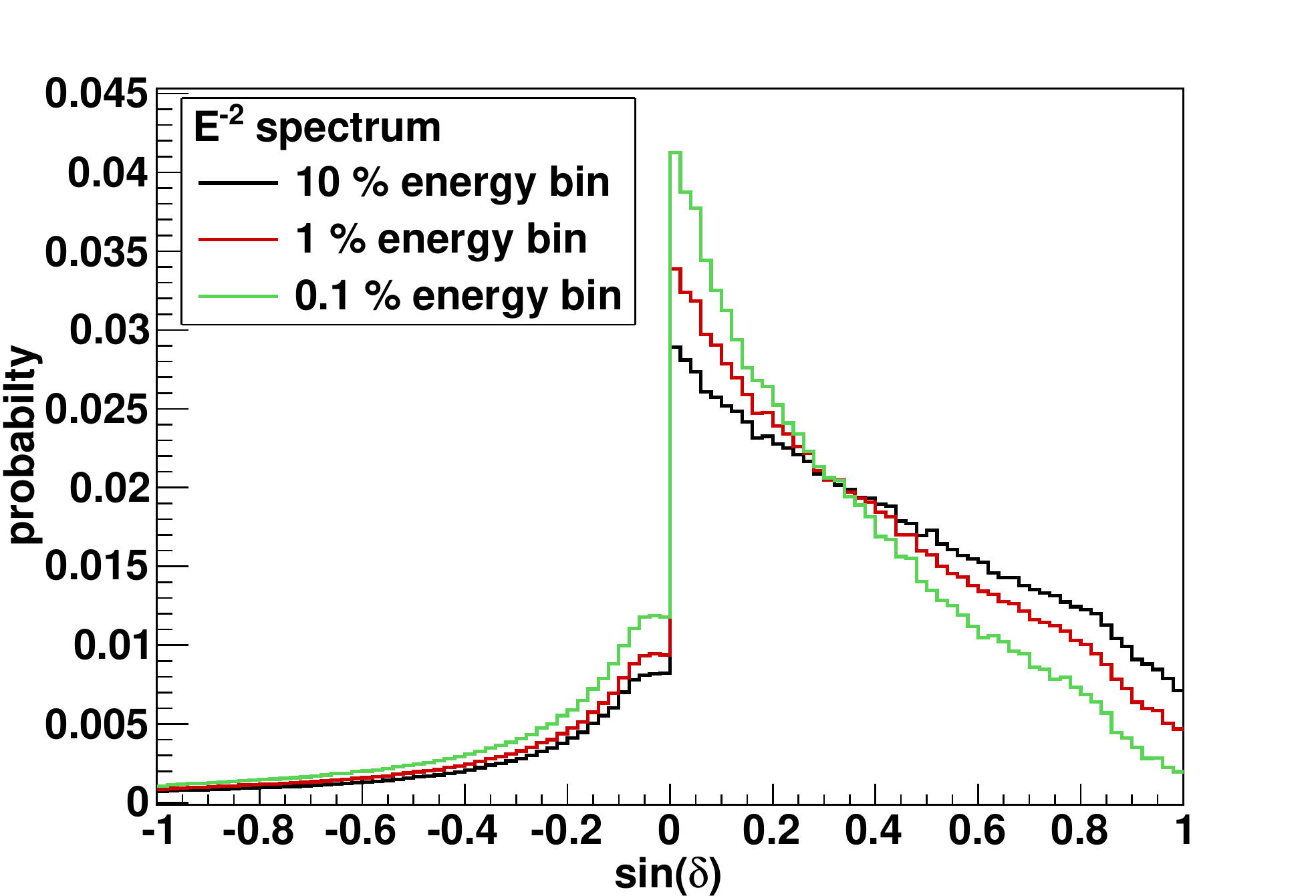} \label{fig:binsZenith} } 
 \caption{Declination distribution $\delta$ for (a) different energy spectra and (b) different energy bins of an $E^{-2}$ spectrum used as detector acceptance for the signal simulation. All histograms are obtained from MC simulation of isotropic signal events.}
 \label{zenithSignal}
\end{figure*}

The declination-dependent detector acceptance, as obtained from MC simulation, is shown in Fig.~\ref{fig:sigZenith} for $E^{-2}$, $E^{-2.25}$ and $E^{-3}$ energy spectra.
Additionally, the detector acceptance for the top 10\%, 1\% and 0.1\% high-energy events of an $E^{-2}$ energy spectrum is shown in Fig.~\ref{fig:binsZenith}. \\
For the northern hemisphere, the expected neutrino flux from the pole region decreases at high energies which is due to the Earth's declining transparency to neutrinos. Thus, for hard energy spectra and the highest-energy bins the signal is dominated by the horizon region, while for soft energy spectra and low-energy bins, the expected number of signal neutrinos from an isotropic signal is largely the same at all declinations. In addition, the declination-acceptance is influenced by declination-dependent cuts, applied to the data sample in~\cite{all_sky}. In the southern hemisphere, the detector acceptance is comparatively much smaller due to strong cuts applied in~\cite{all_sky} to reduce the atmospheric muon background. The observed smooth transition at the horizon is due to the decrease of atmospheric muons and the corresponding increase in signal efficiency.


\subsection{Simulation of Signal and Background Neutrino Sky Maps} \label{sec:SimulationSigBg}	

Using toy MC simulations, pure atmospheric background sky 
maps, sky maps of pure point source signal and mixed sky maps are 
generated. These pseudo-experiments are used to develop the analysis 
method and to quantify the analysis performance (discovery potential). The generation of signal and background neutrinos is described in the following. 

For all maps, the total number of events, $n_{\mathrm{tot}}$, is 
fixed to the number of events in the experimental dataset.
For the signal simulations, three signal 
parameters are used: the mean number of neutrinos per 
source ($\mu$), 
the number of sources ($N_{\mathrm{sou}}$) in the sky (for the corresponding hemisphere) 
and the spectral index ($\gamma$) of the signal energy spectrum. 
The spatial distribution of point sources is assumed to be uniform in 
the full sky. In an additional signal hypothesis, it is assumed to follow the galactic plane.
The number of events produced by the source is drawn 
from a Poisson distribution with a mean source strength of $\mu$.
All source events are distributed 
around the true source position according to the PSF of the specified energy spectrum 
to take the detector's angular resolution into account.
Each generated signal event 
is either rejected from or accepted to the simulated sky map according to its declination using the detector's declination-dependent acceptance in a hit-and-miss procedure~\cite{hitormiss}. 
Thus, $\mu$ is the number of generated neutrinos per source before detector acceptance and corresponds to a flux. The expected number of neutrinos 
per source ($\mu_{\mathrm{eff}}$) in the simulated map depends on the declination, such that $\mu_{\mathrm{eff}}$ is given by the product of $\mu$ and the detector acceptance at the source's declination.
To do this, the MC declination distribution for the specified energy spectrum is normalized such that its peak value is at $1.0$ and taken to be the detector's declination acceptance. Thus, the number of neutrinos from each source within 
one sky map is determined by the declination of the source and Poissonian fluctuations.

The background, consisting of atmospheric neutrinos and 
mis-reconstructed atmospheric muons, is generated by performing a large number of pseudo-experiments using a uniform right ascension
distribution and taking into account the declination distribution from the 
experimental data shown in Fig. \ref{fig:expZenith}. Using this data driven approach, no additional systematic effects 
due to Monte-Carlo simulation are introduced.

\section{Methods}\label{sec:method}

\subsection{Two-Point Autocorrelation Test}\label{sec:method_2pt}

The autocorrelation test is based on the distribution of pairwise 
calculated spatial distances ($\Psi_{ij}$) between events $i$ and $j$, 
which are compared to the background expectation. It is 
designed to detect an event clustering at angular scales $\theta$ comparable 
to the detector resolution, while no prior information about the 
potential sources is required. The amount of clustering is then 
obtained from scanning simultaneously over $\theta$ and different bins for the minimum energy $E_{\mathrm{min}}$ that
optimizes the sensitivity~\cite{finley, hires}.


With $\Psi$ being the spatial distance between two events, the test statistic of the 2-pt analysis can be defined as a function of $\theta$ by
\begin{equation}
TS(\theta, E_{\mathrm{min}}) = 
	\frac{
		\mathrm{obs.~no.~pairs~with~}\Psi_{\mathrm{i},\,\mathrm{j}}\leq\theta,E_{\mathrm{i},\,\mathrm{j}}\geq E_{\mathrm{min}}
	}{
		\mathrm{avg.~no.~bg.~pairs~with~}\Psi_{\mathrm{i},\,\mathrm{j}}\leq\theta,E_{\mathrm{i},\,\mathrm{j}}\geq E_{\mathrm{min}}
	}\, , 
\label{eqn:ts_auto1}
\end{equation}
or, more precisely, by
\begin{equation}
TS(\theta, E_{\mathrm{min}}) = 
	\frac{
		\sum_{i,j\in H, i>j} 
			\Theta(\theta-\Psi_{ij})\cdot\Theta(E_{i,j}-E_{\mathrm{min}})
	}{
		\langle\sum_{m,n\in H, m>n} 
			\Theta(\theta-\Psi_{mn})\cdot\Theta(E_{m,n}-E_{\mathrm{min}})
	\rangle_{bg}}\, ,
\label{eqn:ts_auto2}
\end{equation}
where $E_{i}$ is the reconstructed muon energy of event $i$ (see \cite{EnergyReco} for details on the energy reconstruction), which is a lower bound of the primary neutrino energy and $E_{i,\,j}=\min(E_{\mathrm{i}},\,E_{\mathrm{j}})$. The pairs of events $i$ and $j$ inside the hemisphere, $H$, are counted, with $\Theta$ being the Heaviside function. The background expectation enters in the denominator and is obtained by averaging over a large number of pseudo-experiments using a uniform right ascencion distribution. $E_{\mathrm{min}}$ is the minimum energy defining each energy bin. In the data sample of the northern hemisphere, events have neutrino energies between 240\,GeV and 1.2\,PeV. For the $10\%$ sample, the lower energy bound increases to 6.3\,TeV, while for the $1\%$ and $0.1\%$ samples the lower threshold is 31.6\,TeV and 120\,TeV, respectively \cite{EnergyReco}. For the southern hemisphere, the energies range from 490\,GeV to 8\,PeV. For the $10\%$ sample this range increases to 630\,TeV and for the $1\%$ and $0.1\%$ samples the lower threshold lies at 1.3\,PeV and 2.7\,PeV.
This search is an extension of the Multi Point Source (MPS) analysis that has previously been applied to the Icecube data~\cite{Yolanda}.


\subsection{ Multipole Analyses}\label{sec:method_mp}
A multipole analysis is performed as a second test. It is 
based on the expansion of the measured skymap into spherical 
harmonics which are given by
\begin{equation}
Y_{\ell}^{m}(\vartheta,\,\varphi) = 
    \sqrt{\frac{(2\ell+1)(\ell-m)!}{4\pi(\ell+m)!}} 
    P_{\ell}^{m}(\cos(\vartheta))\exp(\mathrm{i}m\varphi)\, ,
\end{equation}
with $\ell =0,1,2,...$,  $-\ell \leq m \leq \ell$ and 
$P_{\ell}^{m}(x)$ being the Legendre Polynomials. Here,  
$0 \leq \vartheta < \pi$ and $0 \leq 
\varphi < 2 \pi $ are two free parameters of each spherical harmonic 
that can be connected to declination($\delta$) and right ascension($\alpha$) 
by $\varphi = \alpha$ and $\vartheta = \frac{\pi}{2}-\delta$. 

Since spherical harmonics form a complete and orthonormal system, 
any square-integrable function $f(\vartheta,\,\varphi)$ on a sphere 
can be expanded into them, such that it is expressed by a 
superposition of spherical harmonics and the corresponding complex 
expansion coefficients $a_{\ell}^{m}$. The expansion coefficients 
can be obtained by solving the integral
\begin{equation} \label{eq:Expansion}
a_{\ell}^{m} = 
    \int_{0}^{2\pi} \mathrm{d}\varphi 
    \int_{0}^{\pi}\mathrm{d}\vartheta  
        \sin(\vartheta) 
        Y_{\ell}^{*m}(\vartheta,\varphi) 
        f(\vartheta, \varphi)\, .
\end{equation}
In this analysis, $f(\vartheta,\,\varphi)$ is the neutrino arrival 
direction skymap, described by 
\begin{equation}
    f(\vartheta,\,\varphi) = 
        \sum_{i=0}^{n_\mathrm{tot}} 
            \delta^D(\cos(\vartheta)-\cos(\frac{\pi}{2} - \delta_i))\cdot
            \delta^D(\varphi-\alpha_i)
    \, ,
\end{equation}
where $(\delta_i,\,\alpha_i)$ are the delination and right ascension coordinates of event $i$ 
and $\delta^D$ is the Dirac-Delta distribution.
 
 

The expansion is performed using the \textit
{HEALPix} software package\,\cite{healpix}, which splits the sky into 
786\,432 equal sized bins corresponding to an angular size of 
$\sim 0.13^{\circ}$(mean bin radius), which is sufficiently smaller than 
the angular resolution of the events. 

\begin{figure}[tb]
     \centering
     \includegraphics[width=0.6\textwidth]{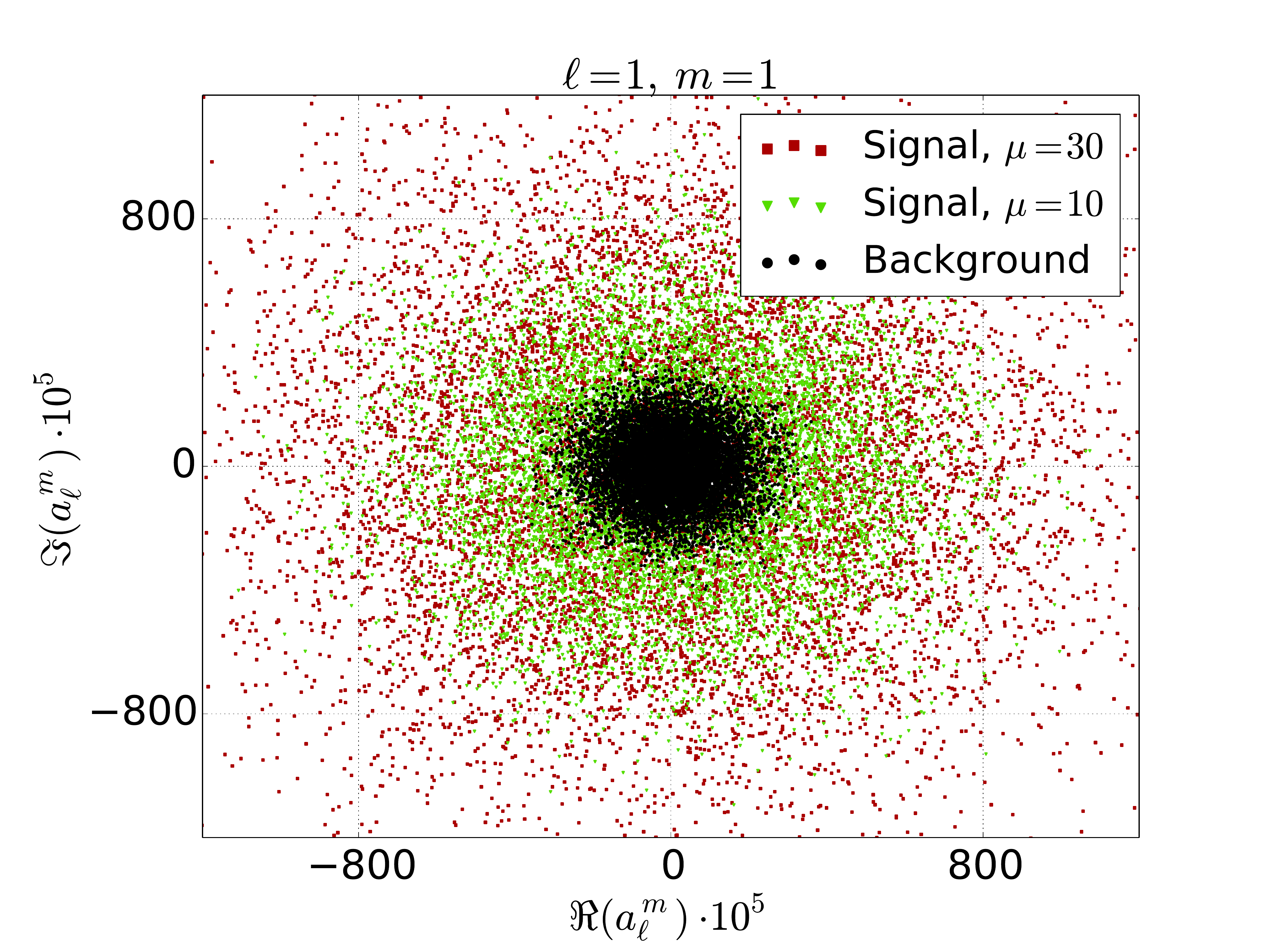}
     \caption{Example of the distribution of a complex expansion coefficient, $a_{\ell=1}^{m=1}$, shown in the complex plane for 10\,000 simulated pure background and pure signal pseudo experiments. For pure signal, two cases of $\mu=10$ and $\mu=30$ are shown. All distributions follow a two-dimensional Gaussian centered at the origin, such that all separation power is contained in the absoulte value of $a_{\ell=1}^{m=1}$, but none in the phase. This is analogous for all other expansion coefficients with $m \neq 0$.}
     \label{fig:almComplPlane}
\end{figure}

For each simulated skymap, the resulting expansion coefficients with $m\neq 0$ are complex quantities where the distribution follows a two-dimensional Gaussian centered at the origin of the complex plane. For different simulated signal strengths, $\mu$, different standard deviations $\sigma$ of the two-dimensional Gaussian are observed, 
while $\sigma$ is minimal for pure background. As an example, the complex expansion coefficient for $\ell=m=1$ is shown in Fig.~\ref{fig:almComplPlane} for 10\,000 simulated sky maps of pure background and pure signal of different source strengths $\mu$. Since the distributions are rotationally symmetric about the origin, no separation power exists within the phase of the expansion coefficients $\phi(a_\ell^m)$, while all separation power rather is contained in their absoulte value $| a_\ell^m |$. \\
The coefficients for $m=0$ describe the pure zenith-dependence and are real numbers. They contain all power 
in the background-only case and thus are not centered at the origin of the complex plane.
Due to its location at the geographic
South Pole and the daily rotation of the earth, 
possible right ascension systematics due to the detector's geometry are averaged out. Remaining possible deviations between the simulated and true signal event distributions are mainly contained in the 
zenith, and thus declination, spectrum of the measured signal. 
Therefore, the $m=0$ coefficients carry almost all of this systematic uncertainty.

\begin{figure}[tb]
     \centering
     \includegraphics[width=0.8\textwidth]{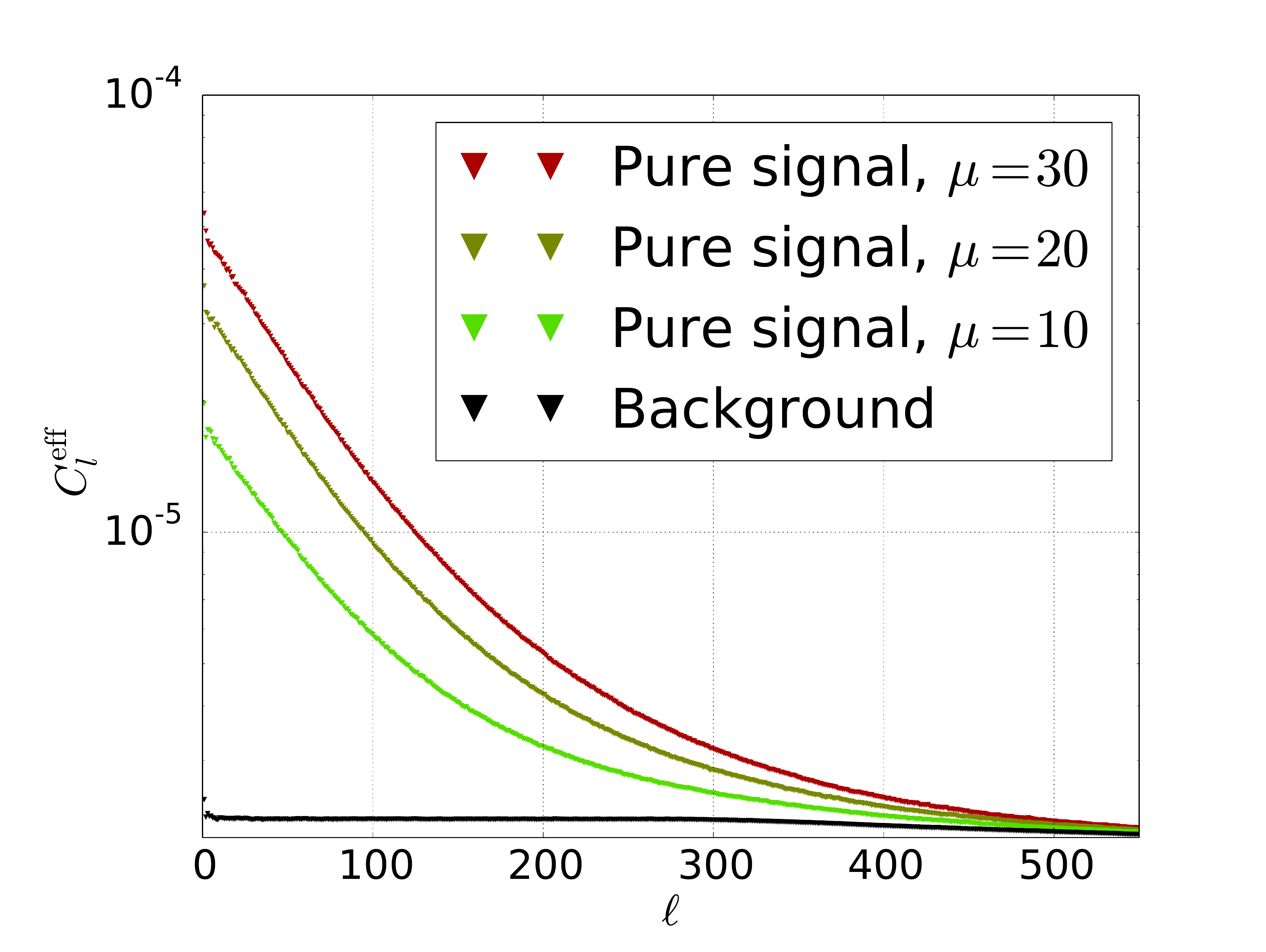}
     \caption{Effective power spectrum $C_{\ell}^{\mathrm{eff}}$ shown for pure signal sky maps and 
 for various values of $\mu$ for an $E^{-2}$ energy spectrum. As described in section~\ref{sec:SimulationSigBg}, for pure signal the number of sources $N_\mathrm{Sou}$ is chosen such that the map contains as many neutrinos as in the experimental sample. The plot shows the averaged values for 10\,000 simulated sky maps.}
     \label{fig:effClSpec}
\end{figure}

Since the complex phase of the expansion coefficients ($\phi$) is 
uniformly distributed, and thus carries no separation power between 
signal and background, the analysis is based on an effective power 
spectrum ($C_{\ell}^{\mathrm{eff}}$) defined by
\begin{equation} \label{eq:EffClDefinition}
C_{\ell}^{\mathrm{eff}} = 
    \frac{1}{2\ell} 
    \sum_{\substack{m=-\ell \\ m\neq 0}}^{\ell} 
        |a_{\ell}^{m}|^{2}\, ,
\end{equation}
using only the absolute values of $ a_{\ell}^{m} $. Additionally, the 
$m=0$ coefficients are omitted due to their different behavior and their strong systematic dependencies. Since at large $\ell$ of $\sim 100$, their contribution to $f(\vartheta,\,\varphi)$  becomes very small, the corresponding loss in performance is negligible. 
The resulting effective power spectra for pure 
signal and pure background averaged over 10\,000 pseudo experiments are shown in Fig. \ref
{fig:effClSpec} .

\begin{figure}[tb]
     \centering
     \includegraphics[width=0.8\textwidth]{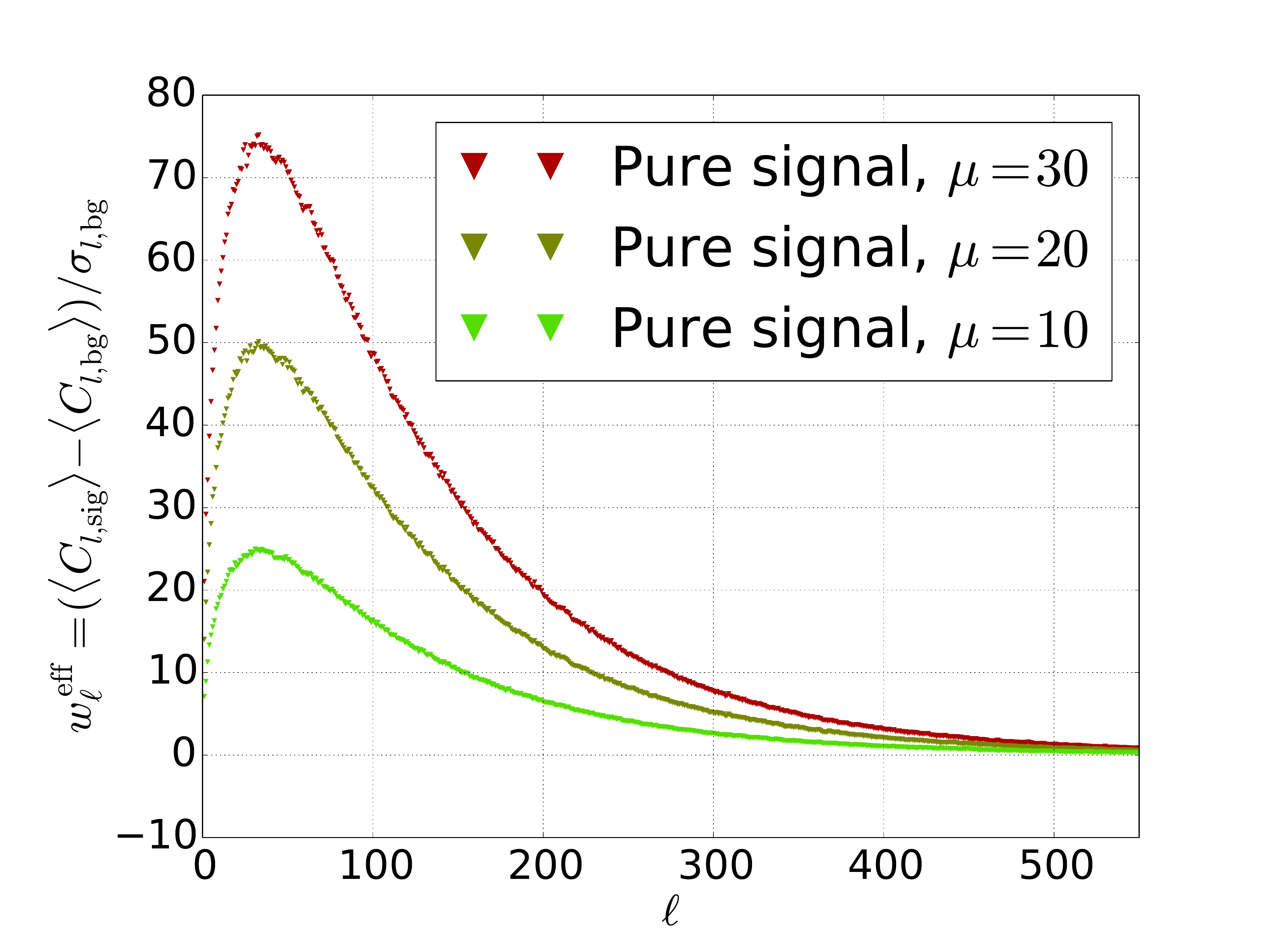}
     \caption{Weight spectrum $w_{\ell}$ shown for pure signal sky maps and 
 for various values of $\mu$ for an $E^{-2}$ energy spectrum. As described in section~\ref{sec:SimulationSigBg}, for pure signal the number of sources $N_\mathrm{Sou}$ is chosen such that the map contains as many neutrinos as in the experimental sample. The weights are calculated from averaged values of $C_{\ell}^{\mathrm{eff}}$ for 10\,000 simulated sky maps. The curves
 exhibit similar shapes and differ only by their 
 total normalization.}
     \label{fig:EffWeights}
\end{figure}


To quantify the differences in the power spectra, a test statistic 
$D_{\mathrm{eff}}^{2}$ is defined by
\begin{equation} 
    \label{eq:TestStatistic}
    D_{\mathrm{eff}}^{2} = 
        \frac{1}{\sum \limits_{\ell=1}^{\ell_{\mathrm{max}}}w^{\mathrm{eff}}_{\ell}} 
        \sum \limits_{\ell=1}^{\ell_{\mathrm{max}}} 
            \left( w^{\mathrm{eff}}_{\ell} \cdot 
            \mathrm{sign}_{\ell} 
            \frac{\left( 
                C_{\ell, \mathrm{exp}}^{\mathrm{eff}} - 
                \langle C_{\ell, \mathrm{bg}}^{\mathrm{eff}}  \rangle
            \right)^{2} }{
                \sigma_{C_{\ell, \mathrm{bg}}^{\mathrm{eff}}}^{2}} \right) 
        \,\mathrm{,}
\end{equation}

which is motivated by a single-sided, weighted $\chi^{2}$-test of 
the experimentally observed, effective power spectrum 
$C_{\ell,\mathrm{exp}}^{\mathrm{eff}}$. The spectra 
$\langle C_{\ell,\mathrm{sig}}^{\mathrm{eff}}\rangle$ and 
$\langle C_{\ell,\mathrm{bg}}^{\mathrm{eff}}\rangle$ are the mean values of the effective 
power spectrum for each $\ell$ for pure signal and pure background sky maps, 
respectively. For all $\ell$,  they are both averaged over 10\,000 pseudo 
experiments. For the pure background sky maps, the corresponding standard deviation for all $\ell$ of the effective power spectrum is called $\sigma_{C_{\ell, \mathrm{bg}}^{\mathrm{eff}}}$.
The parameters $w_{\ell}$ and $\mathrm{sign}_{\ell}$ 
are defined by

\begin{align} 
    \label{eq:Weights}
    w^{\mathrm{eff}}_{\ell} & 
        = \frac{ 
            \langle C_{\ell, \mathrm{sig}}^{\mathrm{eff}} \rangle - 
            \langle C_{\ell, \mathrm{bg}}^{\mathrm{eff}}  \rangle 
        }{
            \sigma_{C_{\ell, \mathrm{bg}}^{\mathrm{eff}}}} \\
    \mathrm{sign}_{\ell}  & 
        = \frac{ 
            C_{\ell, \mathrm{exp}}^{\mathrm{eff}} - 
            \langle C_{\ell, \mathrm{bg}}^{\mathrm{eff}}  \rangle 
        }{
            | C_{\ell, \mathrm{exp}}^{\mathrm{eff}} - 
            \langle C_{\ell, \mathrm{bg}}^{\mathrm{eff}}  \rangle |} 
    \,\mathrm{,}
\end{align}

such that each deviation in $C_{\ell}^{\mathrm{eff}}$ is weighted by 
the expected deviation in the case of a point source signal. Thus,  $C_\ell^{\mathrm{eff}}$ that are very sensitive to point-source signals obtain a large weight, while insensitive $C_\ell^{\mathrm{eff}}$ obtain a small weight to increase sensitivity by keeping the test statistic from being dominated by statistical fluctuations on other angular scales than those relevant for point-source searches. 

Additionally, the parameter $\mathrm{sign}_{\ell}$ guarantees that only deviations in the 
expected direction are counted positively, while deviations in the opposite direction are counted negatively. This is a natural definition, because underfluctuations on scales, where point-sources were supposed to cause an excess, are not meant to count positively for the test statistic, since an underfluctuation does rather contradict a point-source signal on this angular scale than support it.

In Fig. \ref{fig:EffWeights}, the weight spectrum $w_{\ell}$ is shown for various 
source strengths $\mu$ and pure signal. For all $\mu$, it exhibits the similar shape, 
while the shown spectra differ only in their total normalization. Since the test statistic is divided by the sum over all weights (Eq. \ref{eq:TestStatistic}), the test statistic does not depend on the total normalization of the weight spectrum, but only on its shape. Thus, the shown weight spectra lead to the same test statistic. 

A softer spectral index, $\gamma$, would lead to a slightly poorer angular 
resolution. This change in angular resolution also alters the 
characteristic scale of the point sources. Since the weights are 
large for $\ell$ corresponding to the characteristic scale of the structure size, a slightly 
broader angular resolution would lead to a slight shift of the distribution
to lower $\ell$, although this effect and the corresponding loss in sensitivity are small and thus are 
neglected in this analysis. Therefore, in the following only one set of weights, which was calculated for $\mu=30$ and $\gamma=2$,
is applied to experimental data. 


\section{Analysis Performance}\label{sec:DataAnalysis}

\subsection{Discovery Potential of the Autocorrelation Analysis}\label{sec:DisPot2Pt}

In this test, the angular scale $\theta$ is varied from 0$^\circ$ to 5$^\circ$ with a step size of 0.25$^\circ$. In addition, four energy 
bins are used, that contain: all events, the 10\,\%, 1\,\% and 0.1\,\% of the most energetic events observed in data, in order to have 
a better background suppression. The 0.1\,\% sample for the northern hemisphere, for example, contains only the 100 highest-energy events. By using different energy 
thresholds, the discovery potential for high energy signals is 
improved, while the sensitivity to sources with soft energy spectra 
is retained. By varying the step size, the scan itself can 
determine the best energy and $\theta$ binnings that maximize the signal 
\cite{finley}.

The autocorrelation analysis is performed on the data described in section~
\ref{datasample}. The test statistic is acquired from over 20\,000 
pseudo-experiments of randomized data. The test statistic is then fit with a Gamma distribution for the two highest energy 
bins, which contain the 1\,\% and 0.1\,\% of the events with the highest 
energy and a Gaussian for the two lowest energy bins, 
containing all events and the 10\,\% of events with the highest 
energy.
\begin{figure}[tb]
     \centering
     \includegraphics[width=0.7\textwidth]{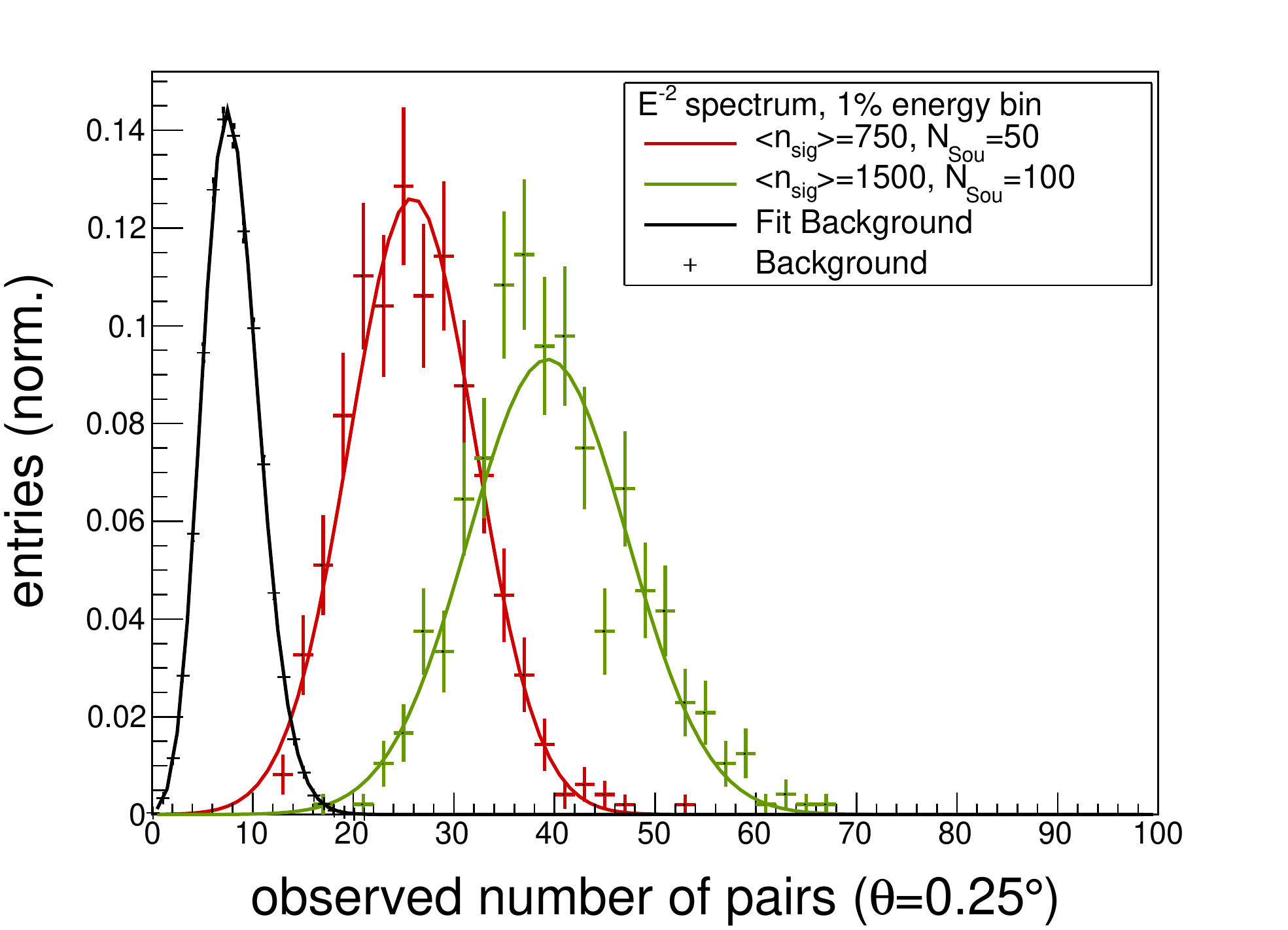} 
     \caption{Example of the number of observed pairs in the autocorrelation analysis for simulated data sets with and without signal. The case for the angular scale $\theta=0.25^\circ$ and the energy bin that contains the 1\,\% highest energy events is shown. The randomized data is fitted with a Gaussian 
     and two signal scenarios with a uniform distribution 
     of $E^{-2}$ sources in the northern sky are shown. For $N_{\mathrm{Sou}} = 50$, the mean number of neutrinos per source $\mu$ was adjusted until the total number of signal events added to the full data sample (and replacing randomized data events at the corresponding declinations) was $n_{\mathrm{sig}}=750$.  Similarly for $N_\mathrm{Sou} = 100$, the example shown was constructed by adjusting $\mu$ per source until the number of signal events inserted in the full data sample was $n_{\mathrm{sig}}=1500$}
     \label{ts_anisotropy}
\end{figure}
Figure \ref{ts_anisotropy} shows the distribution of the number of observed pairs for pure background sky maps, a clustering scale of $0.25^\circ$ and the energy bin containing 1\,\% of the data with a Gaussian fit.
Additionally, two signal scenarios with a uniform distribution of sources in 
the northern sky are shown. 
In order to estimate the discovery potential, the number of signal events 
is varied for different numbers of sources and is compared to the 
test statistic distribution for background. Furthermore, the 
thresholds for the energy bins are re-calculated in order to keep 
the number of events fixed in every energy bin. Given the test statistic distribution for different numbers of signal 
neutrinos ($n_{\mathrm{sig}}$), the required number of signal neutrinos, 
where 50\,\% of the cases give at least a $5\,\sigma $ deviation from the observed background, is defined to be the discovery potential. Applying the solid-angle-averaged effective area 
($A_{\mathrm{eff}}(E)$ - see section~\ref{sec:datasample_chara}) of the 
detector for the three-years (one-year) sample, 
the resulting discovery potential in terms of $n_{\mathrm{sig}}$ can be converted to fluxes. This is done using 
\begin{equation} 
    \label{eq:Numbers2Fluxes}
        E^{\gamma}\frac{\mathrm{d}\phi}{\mathrm{d}E} =  
            \frac{n_{\mathrm{sig}}}{ T_{\mathrm{up}} 
            \int\limits_{0}^{\infty} \mathrm{d}E 
            \, A_{\mathrm{eff}}(E)E^{-\gamma} } 
    \,\mathrm{,}
\end{equation}
where $n_{\mathrm{sig}}$ is the number of signal neutrinos fixed by 
the signal parameters $(N_{\mathrm{Sou}},\, \mu, \, \gamma)$ and 
$T_{\mathrm{up}}$ is the detector uptime.\\
The discovery potential includes the correction for trial factors that come from testing different angular scales $\theta$ and different energy thresholds $E_\mathrm{min}$.
 The first signal model considered here contains sources 
with equal strength and a uniform distribution in the northern 
sky. 
The second scenario is a spatial distribution according to the Green 
catalogue~\cite{Green} of SNRs in the Milky Way, and contains 274 SNRs. Thus, sources are distributed according to randomly chosen positions of SNRs of the catalogue and can vary from 50 to 200 sources. Even if this catalogue gives a biased view of Galactic SNRs, we decided to use it for limit calculations in our search. However, for future applications we will consider improved catalogues of galactic objects. Since the sources are distributed only inside the Galactic Plane, this scenario exhibits a larger clustering between the sources compared to the first one. In order to study the high energy events in more detail we perform an additional test (2-pt HE) for the 100 most energetic events in the IC79 
data sample. A smaller energy binning in steps of 10 
events is applied up to the most energetic 10 events observed in this sample, 
while the binning in $\theta$ stays the same.

\subsection{Discovery Potential of the Multipole Analysis}

\begin{figure}[tb]
     \centering
     \includegraphics[width=0.7\textwidth]{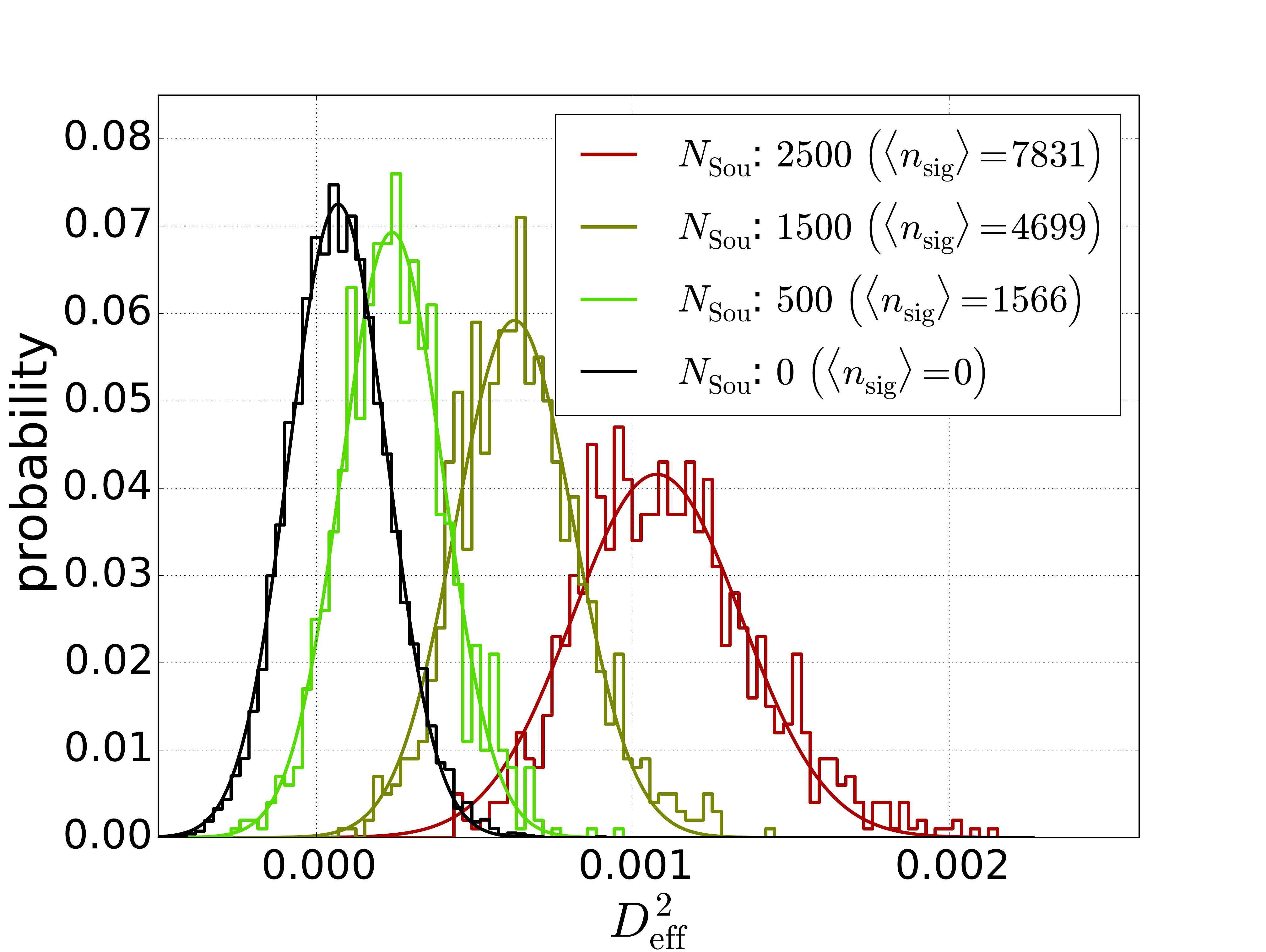}
     \caption{Test statistic for the multipole analysis shown for various amounts of signal 
     in the expanded sky map, given by $\mu=5$, $\gamma=2$ and 
     various values for $N_{\mathrm{Sou}}$. The average number of signal events $n_{\mathrm{sig}}$, which are added to the data in each case, is also shown. A Gaussian fit is shown 
     for all cases and provides good agreement with each 
     distribution.}
     \label{TS_MC}
\end{figure}

The test statistic for mixed sky maps containing only 
a small fraction of signal events and mainly background is calculated to estimate the 
analysis performance. This is done by varying the 
number of sources for fixed values of the mean number of neutrinos per source and 
the spectral index. The resulting test statistic for each 
combination of $(N_{\mathrm{Sou}},\, \mu, \, \gamma)$ is then calculated 
1000 times and 10\,000 times for the background distribution. 
The resulting test statistics for $\mu=5$ are shown in 
Fig. \ref{TS_MC} for an $E^{-2}$ energy spectrum and various 
values of $N_{\mathrm{Sou}}$. To facilitate comparison with the autocorrelation analysis, shown in Fig.~\ref{ts_anisotropy}, the final (average) number of signal events is also given. Note that in contrast to Fig.~\ref{ts_anisotropy}, there is no restriction on the event's energy.
 
The corresponding $5\,\sigma$ discovery potential in terms of the number of signal neutrinos $n_{\mathrm{sig}}$ is defined analogous to section~\ref{sec:DisPot2Pt} and is converted to a physical flux using Eq.~\ref{eq:Numbers2Fluxes}. Further, in contrast to the autocorrelation analysis, the discovery potential of the multipole analysis is determined only for the case of uniformly distributed sources on the northern hemisphere. Furthermore, it does not use a binning in energy or angular distance, such that there is no correction for trials and the resulting pre-trial significance is also the post-trial value. 

\subsection{Comparison of Autocorrelation and Multipole Analyses}
\begin{figure*}[t]
     \centering 
     \subfigure[Discovery potential and limits for the northern sky]{
     \includegraphics[width=1\textwidth]{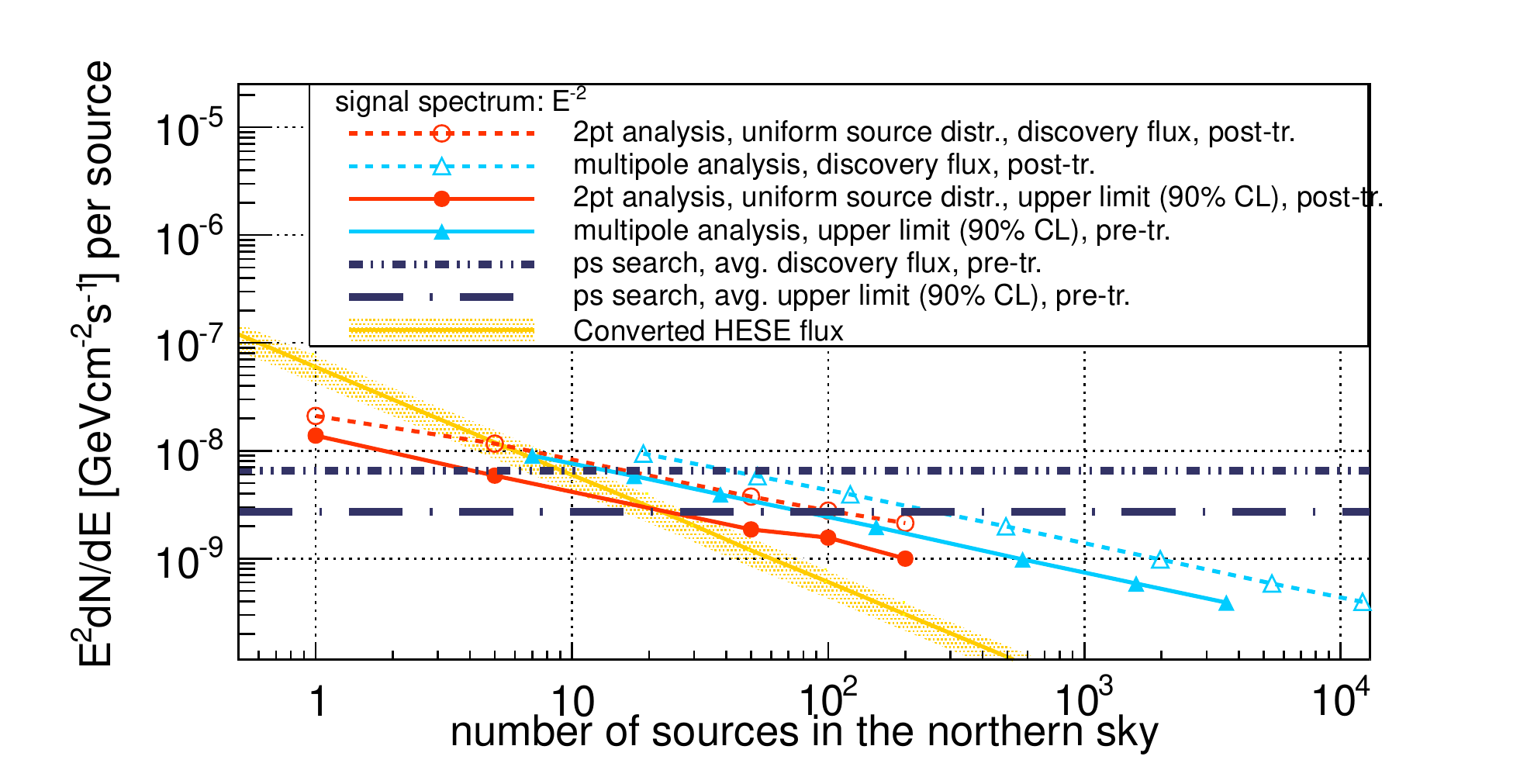}
      \label{limit_north}}
     \subfigure[Discovery potential and limits for the southern sky]{
     \includegraphics[width=1\textwidth]{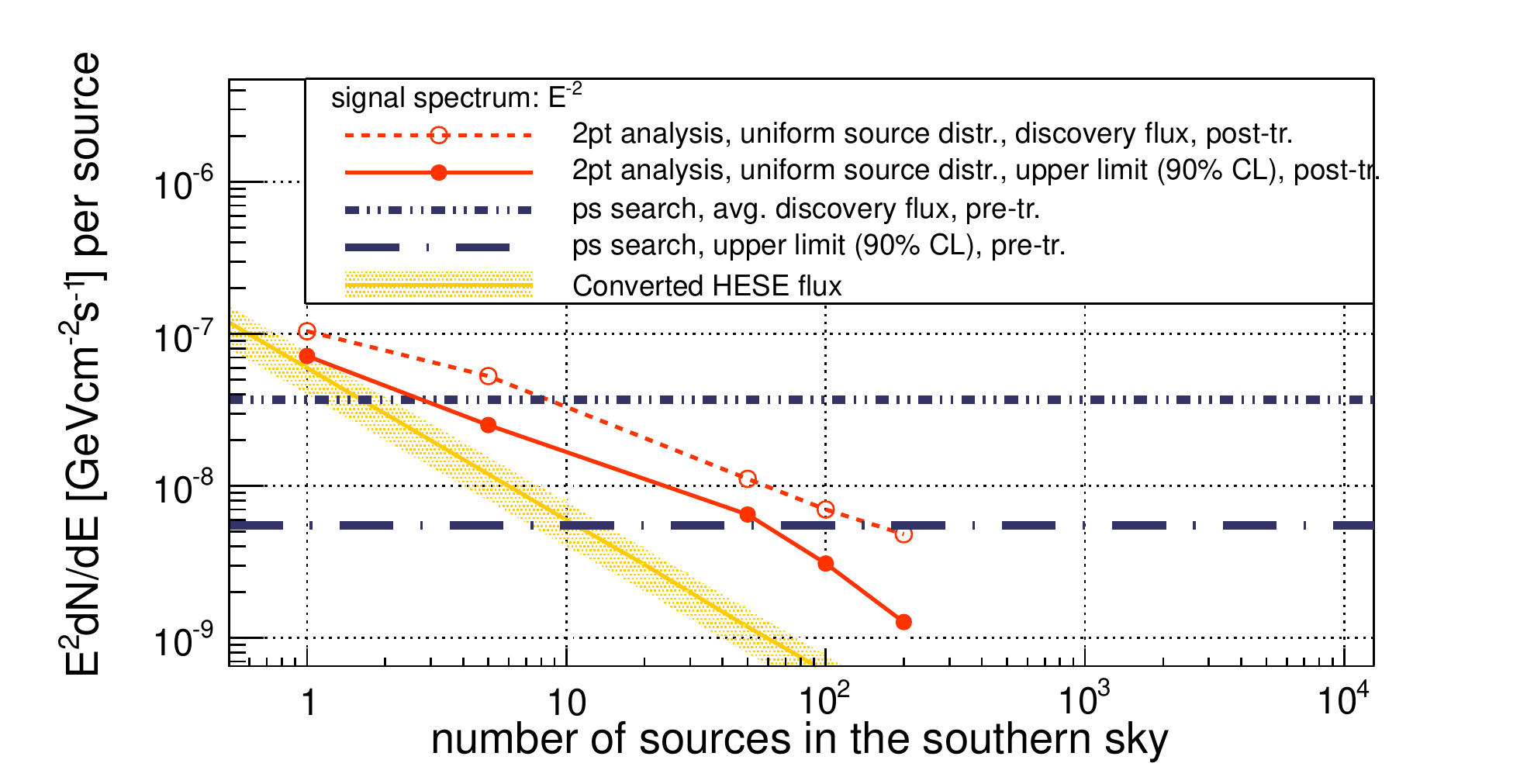}
     \label{limit_south}}
     \caption{Discovery potential and upper limits for uniform $E^{-2}$ neutrino sources for the autocorrelation analysis and the multipole analysis (a) on the northern hemisphere and (b) on the southern hemisphere.
     They are compared to the discovery potential of the point source search~\cite{all_sky}. 
     The yellow band corresponds to the converted flux of the HESE analysis~\cite{HESE}.  
     }
      
\end{figure*}

The $5\,\sigma$ discovery significance for both analyses are shown in Fig.~\ref{limit_north} and Fig.~\ref{limit_south} for the northern and southern hemisphere, respectively, assuming an $E^{-2}$ neutrino spectrum and a uniform source distribution. The significances are compared to the discovery flux of the time-integrated point source likelihood search~\cite{all_sky} averaged for each hemisphere and to the recently found diffuse flux of astrophysical sources~\cite{HESE}. 
This is achieved by evaluating the best fit astrophysical flux for each spectral index which is converted into a flux per source by normalizing with 2$\pi$ for each hemisphere and dividing the diffuse flux by the number of sources, assuming sources of equal flux at Earth.
One should note that the point source discovery flux is shown for a single source and does not include trial factors for searching many locations which, over the whole sky, would increase the discovery flux by a factor of about 2.

\begin{figure*}[t]
     \centering 
     \subfigure[Discovery potential and limits for the northern sky]{
     \includegraphics[width=1\textwidth]{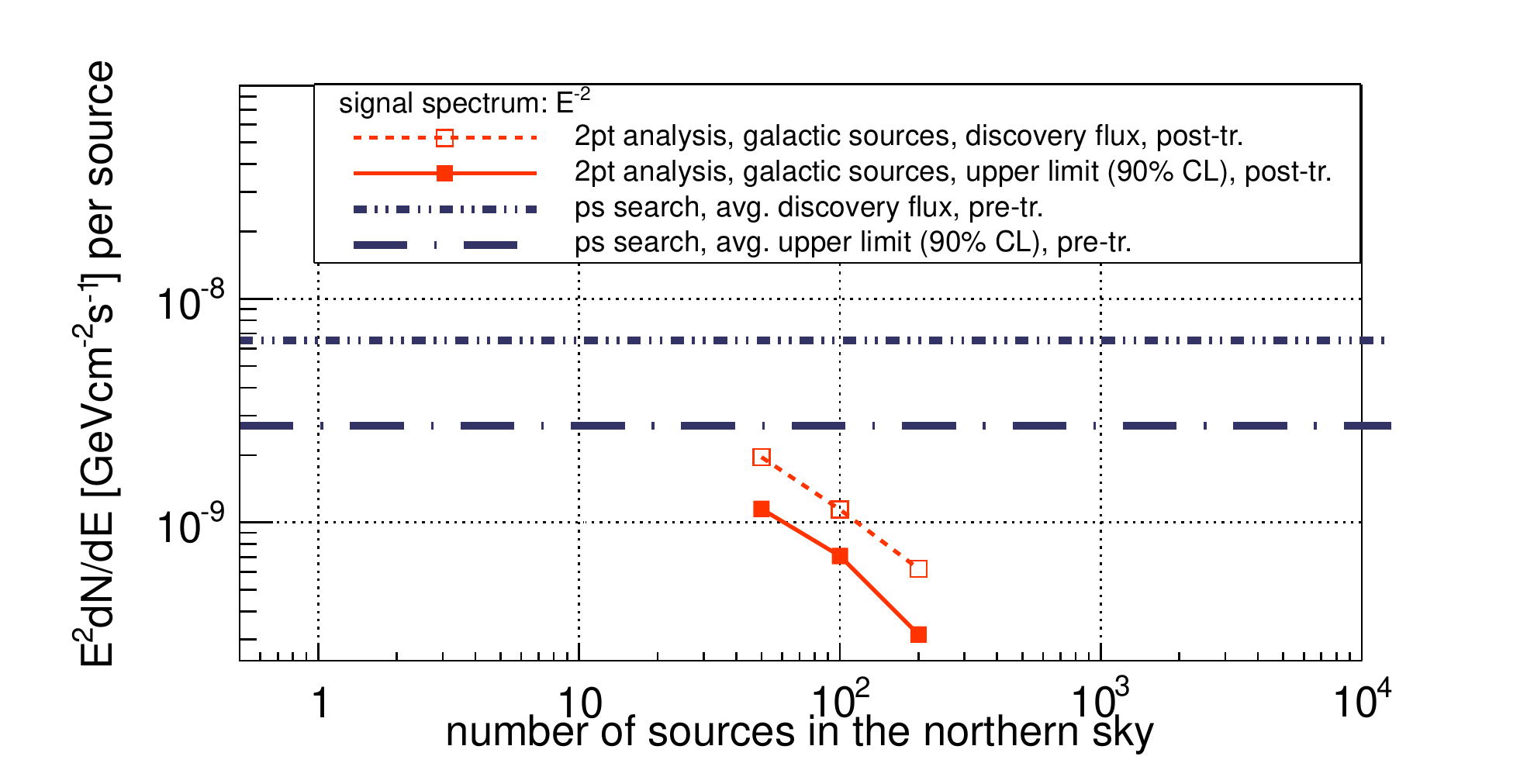}
      \label{limit_north_gp}}
     \subfigure[Discovery potentials and limits for the southern sky]{
     \includegraphics[width=1\textwidth]{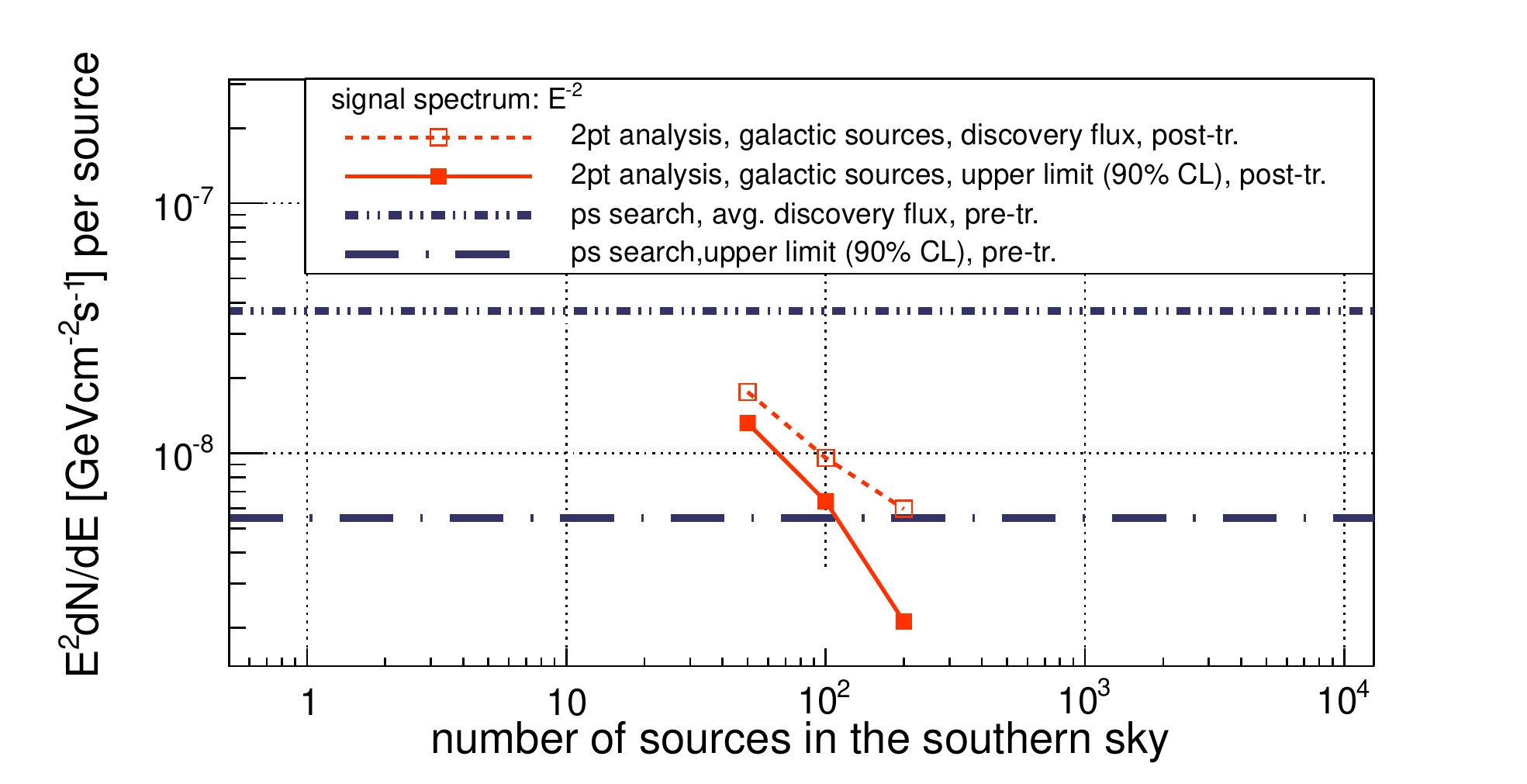}
     \label{limit_south_gp}}
     \caption{(a) The discovery potential and the upper limits for $E^{-2}$ neutrino sources, distributed in the galactic plane with 
     the autocorrelation analysis (a) for the northern hemisphere and (b) for the southern hemisphere. They are compared to the 
     discovery potential of the point source search~\cite{all_sky}. 
   }
\end{figure*}

\begin{figure*}[t]
     \centering 
     \subfigure[Discovery potential and limits for $E^{-3}$]{
        \includegraphics[width=1\textwidth]{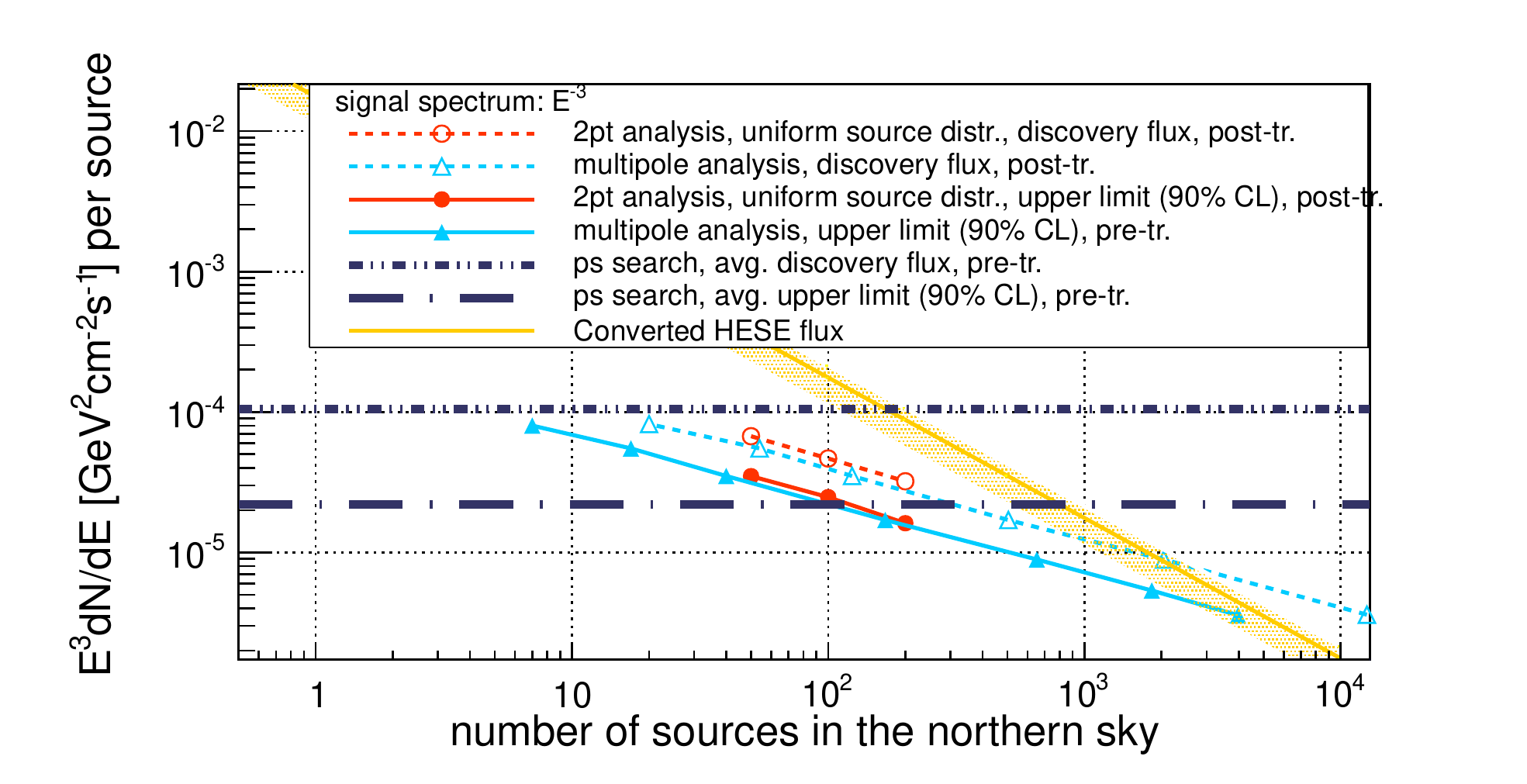}
        \label{limit_e3}
     }
     \subfigure[Discovery potential and limits for $E^{-2.25}$]{
        \includegraphics[width=1\textwidth]{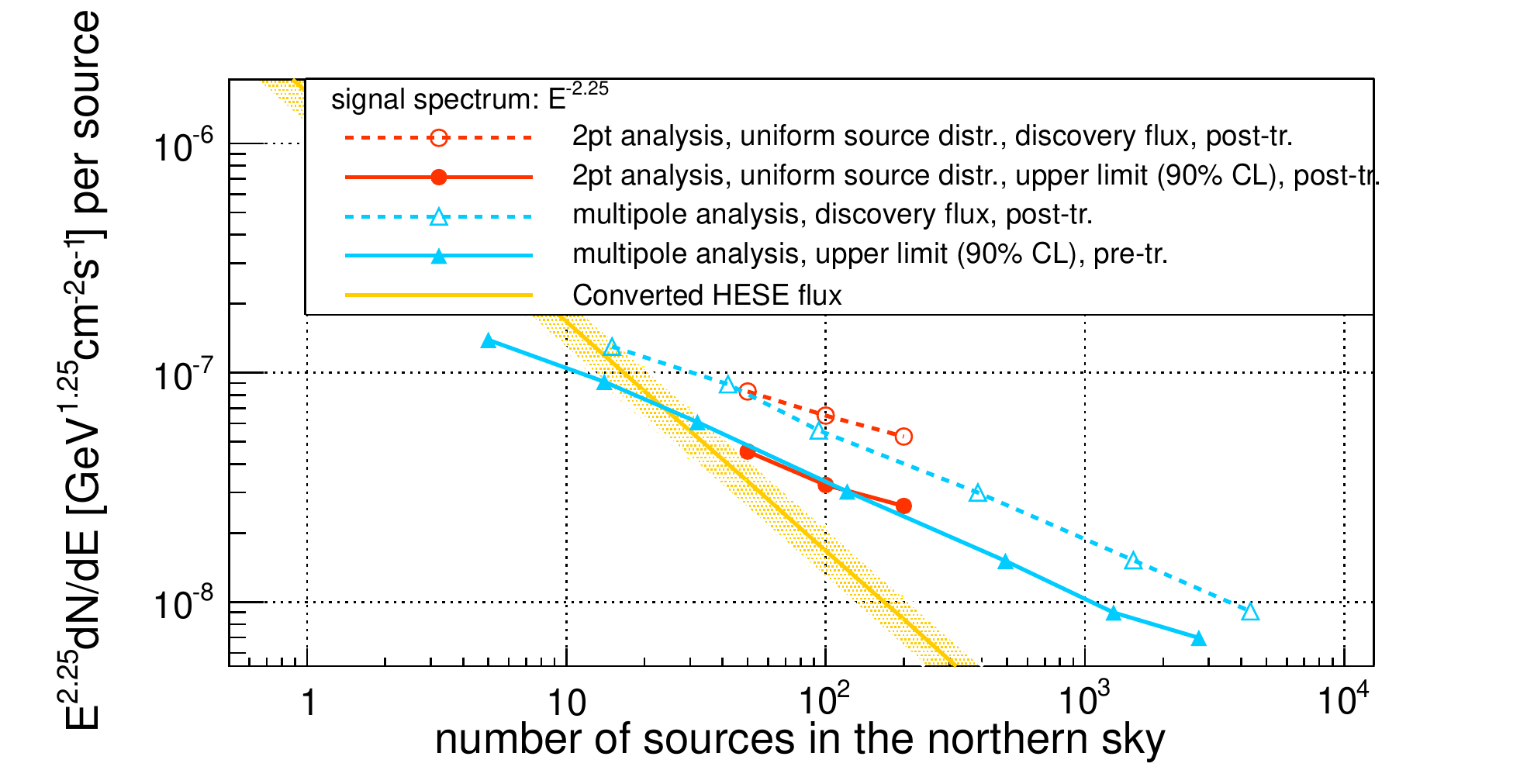}
        \label{limit_e225}
     }
     \caption{(a) The discovery potential and upper limits (a) for $E^{-3}$ neutrino sources and (b) 
     for $E	^{-2.25}$ neutrino sources with 
     the autocorrelation analysis and the multipole analysis for the northern hemisphere. They are compared to the discovery potential and the upper limit of the point source search~\cite{all_sky}. Additionally, the converted flux from the HESE analysis is shown~\cite{HESE}. 
     } 
\end{figure*}

For an $E^{-2}$ spectrum, and more than $\sim 20$ sources in the northern sky, the autocorrelation  analysis is able to identify a signal that the point source likelihood search would not observe, while for the multipole analysis this is the case for more than $\sim 45$ sources. The large difference between these two analyses is due to the hard energy spectrum, which is easier to extract with the autocorrelation analysis since it uses energy as an additional observable. For the southern hemisphere, the autocorrelation analysis is performing better than the point source likelihood analysis for more than $\sim 10 $ sources, while the multipole analysis is not shown, because it was only performed on the northern hemisphere. For the autocorrelation analysis, the $5\,\sigma$ discovery flux  is shown in Fig.~\ref{limit_north_gp} and Fig.~\ref{limit_south_gp} for the northern and southern hemisphere, respectively, assuming an $E^{-2}$ energy spectrum and a galactic source distribution. As described above, it is also compared to the average discovery flux of the point source likelihood search. In Fig.~\ref{limit_e3}, the $5\,\sigma$ discovery flux for an $E^{-3}$ 
neutrino spectrum with uniformly distributed sources is shown and compared to the discovery flux of the point source likelihood search for that spectrum. Above $\sim20$ sources, the multipole and the autocorrelation analysis perform better than the point source likelihood analysis. For this energy spectrum, the discovery potential of both analyses is similar, since the energy observable that is used by the autocorrelation analysis carries only little separation power between astrophysical and atmospheric neutrinos. Figure~\ref{limit_e225} illustrates the discovery flux for sources with an $E^{-2.25}$ neutrino spectrum for uniformly distributed sources. Again, the performance of both analyses is similar, while a discovery potential for the point source likelihood analysis is not available. \\
For a fit of an $E^{-2}$ spectrum to the HESE data, neither analysis is able to detect the underlying source population except in the case that it consists of very few sources (the scenario for a detection in the previous point source likelihood searches).  However, for fits of softer spectral indices to the HESE data like $E^{-3}$ or $E^{-2.25}$, there are a range of source populations compatible with the HESE flux that could be detected by the autocorrelation and multipole analysis, but not by the point source likelihood analysis. For all tested signal hypotheses, the declination-dependent detector acceptance was correctly taken into account. However, it should be noted that in case of a non-isotropic signal, the given discovery potential is not valid since different declinations contribute differently to the total significances (s. Fig.~\ref{zenithSignal}). Additionally, large-scale structures in the source distribution would increase the clustering of events, allowing both methods to detect even smaller fluxes. Thus, the sensitivity depends strongly on the source distribution and can not simply be applied to any model predictions.\\
The calculations presented in this paper are not limited by computation time, although this does scale differently with the number of events in each sky map. The information-based complexity, and thus the required computation time, of the multipole analysis scales with $\mathcal{O}\left(  n_\mathrm{tot}  \right)$, while the autocorrelation analysis scales with  $\mathcal{O}\left(  n_\mathrm{tot}^2  \right)$.


\section{Results}\label{sec:Result}
\subsection{Results of the Autocorrelation Test}

\begin{figure*}[h!]
     \centering
     \includegraphics[width=1\textwidth]{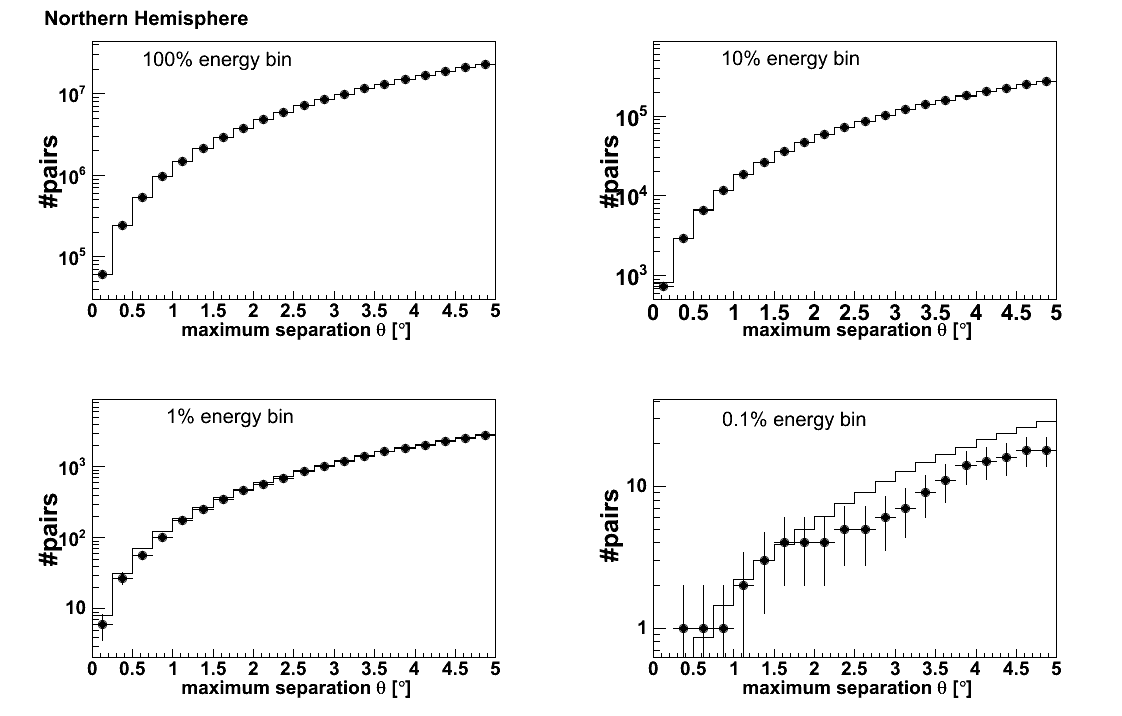}
     \caption{Result for the autocorrelation test on the northern hemisphere as a 
     function of the clustering scale $\theta$. The black points refer 
     to the observed number of pairs, while the black line represents 
     the average number of background pairs. }
     \label{log_north}
\end{figure*}

\begin{figure*}[h!]
     \centering
     \includegraphics[width=1\textwidth]{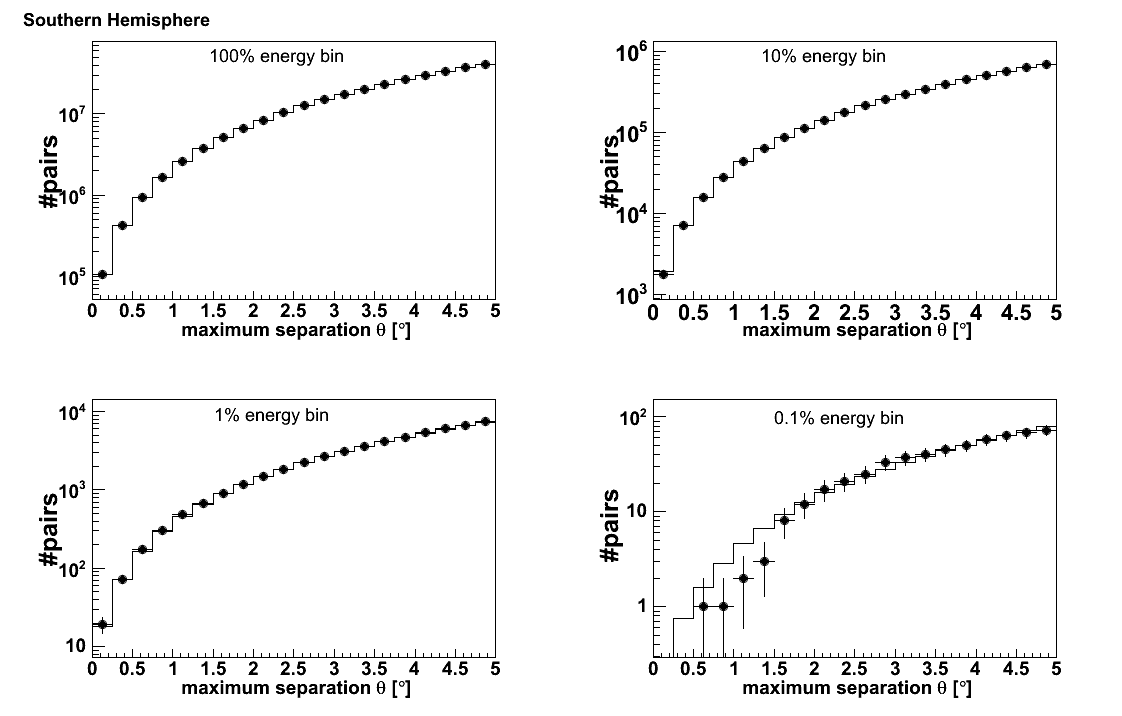}
     \caption{Result for the autocorrelation test for the south as a 
     function of the clustering scale $\theta$. The black points 
     refer to the observed number of pairs, while the black line 
     represents the average number of background pairs.}
     \label{log_south}
\end{figure*}

The autocorrelation test was applied to the presented data sample 
of the IC40, IC59 and IC79 configurations. 
In Fig.~\ref{log_north}, the observed number of pairs for the northern hemisphere are shown as a function of the clustering scale 
$\theta$ and compared to the background expectation. 
The same plots are shown for the southern hemisphere in Fig.~\ref{log_south}. Since the 
data sample contains more events in the southern part of the sky, the fluctuations for the highest energy bin are smaller. 

\begin{figure*}[h!]
     \centering
     \includegraphics[width=1\textwidth]{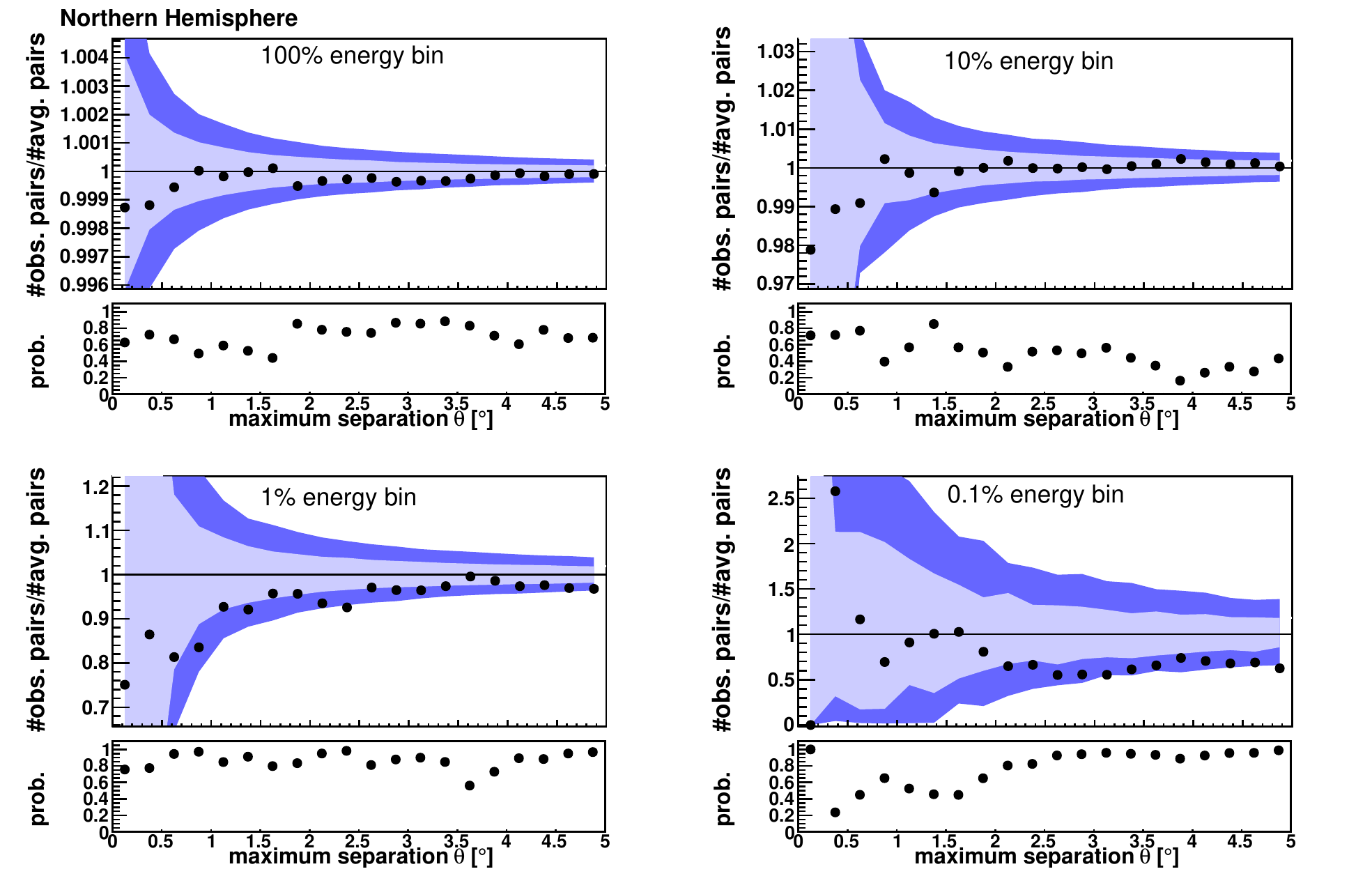}
     \caption{Results for the anisotropy analysis on the northern hemisphere. In each plot, the 
     upper panel shows the ratio of the number of observed pairs and the 
     average number of background pairs, with the +1$\sigma$ and -1
     $\sigma$ (light blue), as well as the +2$\sigma$ and 
     -2$\sigma$ (dark blue) contours as a function of the clustering scale $\theta$. 
     The lower panel shows the probability before trials. The best 
     p-value for the northern hemisphere is 0.16 and is found at $\theta<4^{\circ}$ in the highest-energy 10
     \% selection. The final p-value after correcting for trials is 0.84.}
     \label{result_north}
\end{figure*}

\begin{figure*}[h!]
     \centering
     \includegraphics[width=1\textwidth]{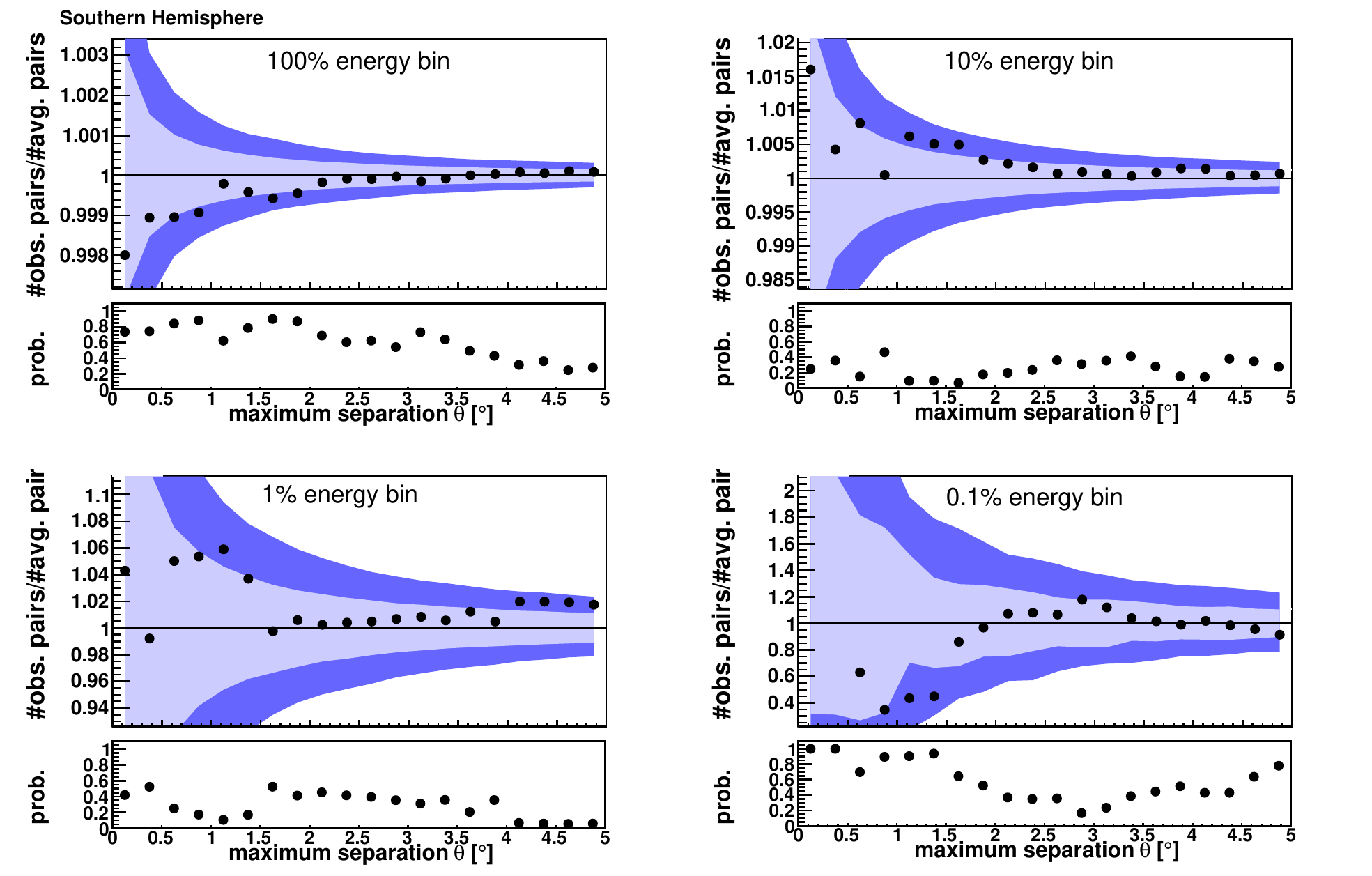}
     \caption{Results for the anisotropy analysis for the southern hemisphere. For each plot, the 
     upper panel shows the ratio of the number of observed pairs and 
     the average number of background pairs, with the +1$\sigma$ and 
     -1$\sigma$ (light blue), as well as the +2$\sigma$ and 
     -2$\sigma$ (dark blue) contours for the clustering scale $\theta$. The 
     lower panel shows the probability before trials. The best 
     p-value for the southern hemisphere is 0.055 and is found at $\theta<4.75^{\circ}$ in the highest-energy 1
     \% selection. The final p-value after correcting for trials is 0.73.}
     \label{result_south}
 \end{figure*}

The ratios of the observed number of pairs and the background 
expectation of this analysis are shown in Fig.~\ref{result_north} 
for the northern hemisphere and Fig.~\ref{result_south} for the 
southern hemisphere. In both hemispheres, fewer pairs than 
expected were observed and a small underfluctuation is visible. The 
background distribution is used for the evaluation of the local 
p-values. The best pre-trial p-values are 0.16 for the northern and 0.055 for the southern hemisphere. Taking the trials for the different angular and energy bins into account results in a post-trial p-value of 0.84 for the northern 
hemisphere and 0.73 for the southern hemisphere. The result of the additional high energy test of IC79 data only (2-pt HE) gives a pre-trial
p-value of 0.035. After taking the trial correction into account results into a post-trial p-value of 0.38, with one pair at 1.75${^\circ}$ inside the 
10 events energy bin. The pair of events has reconstructed (right ascension, 
declination) of (285.7$^\circ$,+3.1$^\circ$) and (287.2$^\circ$,+3.6$^\circ$). All results are consistent 
with fluctuations of the background.

\subsection{Results of the Multipole Analysis}

\begin{figure*}[t]
     \centering
     \includegraphics[width=0.7\textwidth]{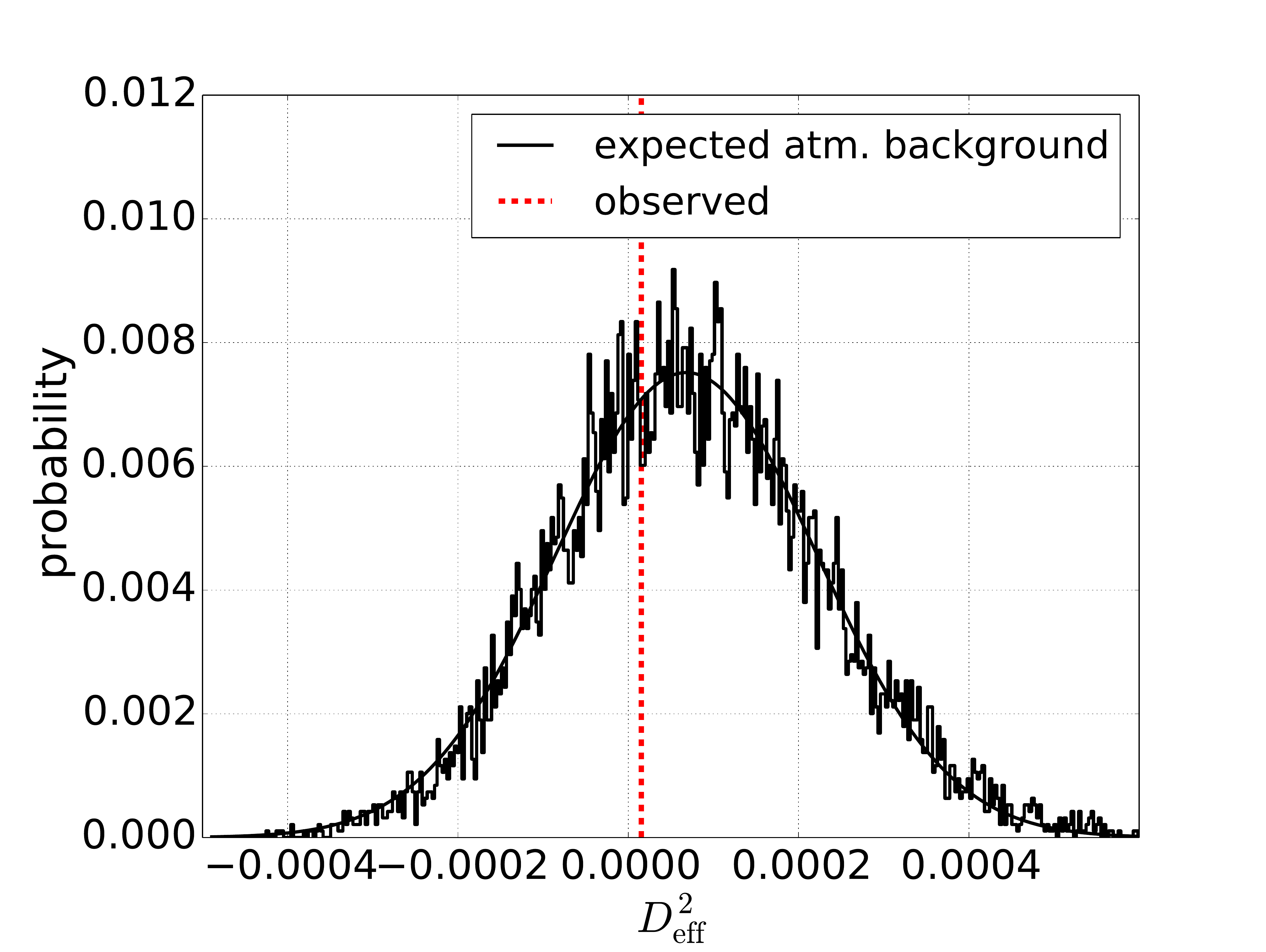}
     \caption{Test statistic for the multipole analysis for pure background compared 
     to the experimentally observed value.}
     \label{fig:TS_EXP}
\end{figure*}

The method is applied to experimental data from the northern hemisphere by calculating 
the experimental power spectrum and the value of the test statistic. 
The observed value of the test statistic is shown in Fig.~\ref
{fig:TS_EXP} with the background expectation. This corresponds to a 
$-0.3\,\sigma$ deviation from background and thus an 
underfluctuation with a p-value of $0.63$ to find a higher value of $D^2_{\mathrm{eff}}$. 

\subsection{Upper Limits on the neutrino flux}

Using Eq.~\ref{eq:Numbers2Fluxes}, the experimental value of the test 
statistic for both analyses can be converted into upper limits on $n_{\mathrm{sig}}$ and thus on the physical flux 
normalizations 
($\Phi_{0}=E^{\gamma}\frac{\mathrm{d}\Phi}{\mathrm{d}E}$) for 
different spectral indices ($\gamma$). Upper limits at the 90\% confidence level are calculated based on the classical (frequentist) approach~\cite{Neyman2}. The resulting 
limits on the flux normalization for the northern hemisphere and uniformly distributed $E^{-2}$ neutrino 
sources are shown in Fig.~\ref{limit_north}. For comparison,
 additional lines are drawn for the limit of the point source likelihood analysis~\cite{all_sky} 
 and for the converted flux of the HESE analysis.
 The average upper limit of the point source likelihood analysis is $2.7 \cdot 10^{-9} \text{GeV}/\text{cm}^{-2}\text{s}^{-1}$. Figure~\ref{limit_south} shows the limit for 
the southern hemisphere and uniformly distributed $E^{-2}$ neutrino sources for the 
autocorrelation analysis. Again, the limit for the point source likelihood analysis and the converted HESE flux are shown for comparison and the average upper limits of the point source likelihood analysis is $5.5 \cdot 10^{-9} \text{GeV}/\text{cm}^{-2}\text{s}^{-1}$.

In Fig.~\ref{limit_north_gp} and Fig.~\ref{limit_south_gp}, the limits are shown for the northern and southern hemisphere for an $E^{-2}$ neutrino spectrum in the galactic plane scenario. Both are compared to the converted HESE flux and the limits from the point source likelihood analysis. 

For both analyses, the choice of sampling points is different due to different approaches in determining the limit by varying the signal parameters $(N_\mathrm{Sou}, \mu, \gamma)$. While the autocorrelation analysis fixes $N_\mathrm{Sou}$ and varies $n_\mathrm{sig}$ by changing $\mu$ to determine the limit, the multipole analysis varies $N_\mathrm{Sou}$ keeping $\mu$ fixed. This results in a different set of sampling points for the limit lines, which has no influence to the lines themselves or their physical interpretation.

In Fig.~\ref{limit_e3} and Fig.~\ref
{limit_e225}, the 
limits of both analyses are shown for an $E^{-3}$ and an $E^{-2.25}$ spectrum, respectively. 
They are compared to the converted HESE flux and to the limit of the point source likelihood analysis for $E^{-3}$, while for $E^{-2.25}$ no limit from the point source likelihood analysis is available.
The $E^{-2.25}$ spectrum is motivated by the HESE best fit of the spectral index.

\section{Systematic Influences}\label{sec:systematics}

The analyses are affected by only few systematic uncertainties due to uncertainties in the background estimation and in 
the signal efficiency. To estimate the systematic influences we use only the IC79 
data sample since it contains most of the events in the 
combined sample and this is a good approximation. 

The background estimation is based on randomized experimental data, and systematic effects in the background can only result from preexisting large-scale anisotropies in the experimental sample. Therefore,
no systematic effects on background estimation are introduced due to MC simulations, \textit{e.g.} from assumptions on any hadronic models or on the 
composition of cosmic rays. 

Since point source searches look for small-scale anisotropies, these must be distinguishable from possible large-scale structures in the data sample. To investigate the effect of a preexisting large-scale
anisotropy in the background, the analyses are repeated including a large-scale anisotropy in the background of atmospheric neutrinos and muons for the mixed sky maps, while keeping the test statistic of the null hypothesis fixed. The large-scale anisotropy is simulated according to a measurement by Milagro~\cite{milagro}. Since the Milagro anisotropy is an anisotropy in cosmic rays, a possible anisotropy of atmospheric muons and neutrinos of about the same order of magnitude is expected. The corresponding systematic errors are given by the resulting shift in the sensitivity of the analyses.


Systematic uncertainties in the signal efficiency arise mainly from the
DOM-efficiency and optical ice properties. In this context the 
DOM-efficiency describes the absolute light detection efficiency of the optical 
modules. The variation of the ice parameters refers to the optical 
properties of the ice, including absorption and 
scattering. 
These systematic effects are estimated using MC simulations analogously to~\cite{all_sky}.
To estimate the systematic errors, the following three uncertainties 
are propagated through the analysis while calculating the 
sensitivity:
\begin{enumerate}
    \item Variation of the DOM-efficiency by $\pm 10\%$
    \item Variation of the Ice Parameters by $\pm 10\%$
    \item Influence of a large-scale anisotropy in the background estimation.
\end{enumerate}

The resulting effects on the sensitivities are shown in Table~\ref 
{tab:Sys}. While the uncertainties for the DOM-efficiency and absorption and 
scattering in the ice are the same for both 
analyses, different values for the uncertainties due to a large-scale anisotropy are shown.

\begin{table}
    \begin{center}
        \begin{tabular}{ c | c | c | c | c | c | c }
			\hline
			\hline
            \multirow{3}{*}{spectrum} & \multicolumn{6}{ c }{ effect on sensitivity }  \\
                                    &  DOM Eff. &  Abs. \& Scatt. &  \multicolumn{2}{ c }{Large-Scale Aniso. } &  \multicolumn{2}{| c }{Combined} \\
            & & & Multip. & 2-pt. & Multip. & 2-pt.\\
            \hline
            $E^{-2}$	            & $\pm 13\%$ & $\pm 6\% $ &$ \pm 6\% $ & $ \pm 4\% $ & $ \pm 16\% $ & $ \pm 15\% $\\
            $E^{-2.25}$	            & $\pm 14\%$ & $\pm 6\% $ & $\pm 8 \% $ & $ \pm 8\% $ & $ \pm 17\% $ & $ \pm 17\% $\\
            $E^{-3}$ 				& $ \pm 26\% $ & $ \pm 11\% $ & $ \pm 3\% $ & $ \pm 6\% $ & $ \pm 28\% $ & $ \pm 29\% $\\
            \hline
            \hline
        \end{tabular}
    \end{center}
    \caption{Systematic errors on the flux normalization for the 
    combined data sample while varying the DOM-efficiency and 
    absorption and scattering of the ice. The uncertainty due to a large-scale 
    anisotropy in the background estimation is calculated separately for each 
    analysis. The maximum uncertainty for both analysis is listed above.}
    \label{tab:Sys}
\end{table}


\section{Conclusions}\label{sec:conclusion}

Two methods to search for a small-scale anisotropy with IceCube 
were presented. 
The results of both searches are consistent with
background expectations with small underfluctuation. 
Depending on the number of assumed sources, the resulting upper limits range from $10^{-8}$ GeV/$\mathrm{cm}^{2}\mathrm{s}^{-1}$ for one source to $10^{-9}$ GeV/$\mathrm{cm}^{2}\mathrm{s}^{-1}$ for 3500 $E^{-2}$ neutrino sources in the northern hemisphere. Limits were also set for other assumed energy spectra, including $E^{-3}$ and $E^{-2.25}$ in the northern hemisphere. Since both analyses 
use a data-driven background estimation they are more robust 
against systematic uncertainties than estimations from MC 
simulations. \\
Considering the astrophysical flux previously observed in IceCube~\cite{HESE}, a small number ($\leq 10$) of isotropically distributed sources in the northern hemisphere of very hard energy spectra, like $E^{-2}$, is excluded as it was by former IceCube analyses~\cite{all_sky}. For softer energy spectra, the analyses presented here disfavor the observed flux to come from less than $\sim20$ sources for $E^{-2.25}$ and from less than $\sim 5000$ sources for $E^{-3}$.
Additionally, for sources distributed along the galactic plane in the northern hemisphere the autocorrelation limit is close to the flux predicted by HESE. In the southern hemisphere, the data sample contains predominantly atmospheric muons from cosmic ray air showers above the detector. Due to this background the autocorrelation analysis is not sensitive to a population of sources at the HESE flux level. For all these tests, the sources are assumed to have the same flux at Earth, since the true spatial flux distribution is not known for the observed astrophysical flux. \\
For hard energy spectra, the 2-pt correlation analysis is more sensitive than the multipole analysis since it uses the energy information as an additional variable. For soft energy spectra, the multipole analysis becomes slightly more competitive. \\

{\begin{small}

 \textbf{Acknowledgements} We acknowledge the support from the following agencies: U.S. National Science Foundation-Office of Polar Programs, U.S. National Science Foundation-Physics Division, University of Wisconsin Alumni Research Foundation, the Grid Laboratory Of Wisconsin (GLOW) grid infrastructure at the University of Wisconsin - Madison, the Open Science Grid (OSG) grid infrastructure; U.S. Department of Energy, and National Energy Research Scientific Computing Center, the Louisiana Optical Network Initiative (LONI) grid computing resources; Natural Sciences and Engineering Research Council of Canada, WestGrid and Compute/Calcul Canada; Swedish Research Council, Swedish Polar Research Secretariat, Swedish National Infrastructure for Computing (SNIC), and Knut and Alice Wallenberg Foundation, Sweden; German Ministry for Education and Research (BMBF), Deutsche Forschungsgemeinschaft (DFG), Helmholtz Alliance for Astroparticle Physics (HAP), Research Department of Plasmas with Complex Interactions (Bochum), Germany; Fund for Scientific Research (FNRS-FWO), FWO Odysseus programme, Flanders Institute to encourage scientific and technological research in industry (IWT), Belgian Federal Science Policy Office (Belspo); University of Oxford, United Kingdom; Marsden Fund, New Zealand; Australian Research Council; Japan Society for Promotion of Science (JSPS); the Swiss National Science Foundation (SNSF), Switzerland; National Research Foundation of Korea (NRF); Danish National Research Foundation, Denmark (DNRF)
\end{small}}\\

\bibliographystyle{elsearticle/model1-num-names}
\bibliography{mybib}

\end{document}